# Structure of the photo-catalytically active surface of SrTiO$_3$

Manuel Plaza,[1,µ] Xin Huang,[µ] J. Y. Peter Ko,[2] and Joel D. Brock[*]
School of Applied and Engineering Physics, Cornell University, Ithaca, New York 14853, USA

Mei Shen, Burton H. Simpson, and Joaquín Rodríguez-López[µ]
Department of Chemistry, University of Illinois at Urbana-Champaign, Urbana, Illinois 61801, USA

Nicole L. Ritzert, and Héctor D. Abruña
Department of Chemistry and Chemical Biology, Cornell University, Ithaca, New York 14853, USA

Kendra Letchworth-Weaver,[µ] Deniz Gunceler, and T. A. Arias
Department of Physics, Cornell University, Ithaca, New York 14843, USA

Darrell G. Schlom
Department of Materials Science and Engineering, Cornell University, Ithaca, New York 14853, USA

[1]Current Address: Departamento de Física de la Materia Condensada, Universidad Autónoma de Madrid, Ciudad Universitaria de Cantoblanco, 28049, Madrid, Spain
[2]Current Address: Cornell High Energy Synchrotron Source, Cornell University, Ithaca, NY 14853, USA
[µ]Equal contribution
[*]Corresponding Author

Correspondence to: jdb20@cornell.edu

**A major goal of energy research is to use visible light to cleave water directly, without an applied voltage, into hydrogen and oxygen. Since the initial reports of the ultraviolet (UV) activity of $TiO_2$ and $SrTiO_3$ in the 1970's, researchers have pursued a fundamental understanding of the mechanistic and molecular-level phenomena involved in photo-catalysis.[1-7] Although it requires UV light, after four decades $SrTiO_3$ is still the gold standard for splitting water. It is chemically stable and catalyzes both the hydrogen and the oxygen reactions without applied bias. While ultrahigh vacuum (UHV) surface science techniques have provided useful insights,[8] we still know relatively little about the structure of electrodes in contact with electrolytes under operating conditions. Here, we report the surface structure evolution of a $SrTiO_3$ electrode during water splitting, before and after training with a positive bias. *Operando* high-energy X-ray reflectivity measurements demonstrate that training the electrode irreversibly reorders the surface. Scanning electrochemical microscopy (SECM) at open circuit correlates this training with a tripling of the activity toward photo-induced water splitting. A novel first-principles joint density-functional theory (JDFT) simulation constrained to the X-ray data via a generalized penalty function identifies an anatase-like structure for the more active, trained *surface*.**

Wide bandgap, *n*-doped, metal oxide semiconductors such as $SrTiO_3$, $TiO_2$ and $WO_3$ absorb UV light to form photo-generated (electrons and holes) charge carriers, capable of driving redox reactions at the interface with an electrolyte.[9-11] $SrTiO_3$ is a prototypic perovskite structure metal oxide (Figure 1a). The perovskites exhibit a vast range of attractive physico-chemical properties including promising energy conversion activity.[12] In particular, $SrTiO_3$ has very attractive photocatalytic properties. It is highly stable in base, displays high quantum efficiency for the electro-oxidation of water under UV illumination, and performs the light-driven water splitting reaction (*i.e.*, photo-generation of both $O_2$ and $H_2$ from $H_2O$) under an applied bias,[2,13] at open circuit aided by an auxiliary metal electrode,[2,13] and on free-standing crystals.[4,5,13-16] The fundamental mechanisms underlying surface photochemical reactions, however, remain unclear. While the bulk *d*-band structure can correlate with activity,[12] surface defects and surface structure are critically important and it is typically difficult to decouple the bulk and surface contributions to observed changes in reactivity.[17-19] Here, we illustrate the critical role that surface structure plays by demonstrating under *operando* conditions that the electrochemical activation (training) of *n*-doped $SrTiO_3$(001) in basic media induces an irreversible surface reordering that enhances (by 260%) its activity for photocatalytic water splitting.

The *operando* structural characterization of $SrTiO_3$ before and after training was performed using high-energy X-ray reflectivity measurements during photocatalytic water splitting.[20,21] Taking advantage of the penetrating power of high energy (30 keV) X-rays, a novel photo-electrochemical cell (Figure 1b), and a large format area detector optimized for hard X-rays, we measured absolute X-ray structure factors. The sub-Angstrom sensitivity of X-ray reflectivity to average surface/interface structure allows us to characterize the $SrTiO_3$(001)/electrolyte interface and the bulk-like layers beneath, while maintaining strict electrochemical reaction conditions.

Scanning Electrochemical Microscopy (SECM)[22] was used to follow sequential functional changes (*e.g.*, $O_2$ evolved or adsorbed intermediate O)[23,24] at the surface. Joint density-functional theory (JDFT)[25,26] was employed to calculate the electronic structure, geometry, and X-ray structure factors of the $SrTiO_3$ surface in the presence of a microscopically detailed liquid environment. A detailed description of each of these methods is given in the Supplementary Information.

We carefully prepared the $SrTiO_3$ electrode to ensure an atomically smooth surface with well-defined doping and benchmarked its UV light-induced water-splitting activity to optimize our experimental design. As shown in Figure 1c, chemical etching produced atomically flat terraces on the $SrTiO_3$(001) surface. Precisely controlled vacuum annealing, followed by a thermal quench, fixed the bulk oxygen vacancy concentration and concomitant carrier density to $n_D \sim 10^{17}$ cm$^{-3}$, producing samples with a clear photocurrent signal in 0.1 M NaOH electrolyte under illumination with a 200 W Hg/Xe lamp. Figure 1d shows a typical photocurrent vs. applied potential profile under chopped illumination, where the anodic photocurrent, corresponding to the electrochemical oxidation of water to form $O_2$, exhibits a plateau at ~ +0.8 V vs. Ag/AgCl. Large photocurrents occur only under UV irradiation. Monochromatic UV light, centered at $\lambda = 390 \pm 20$ nm (~3.2 eV), was used in subsequent experiments to control the illumination conditions precisely.

Figure 2a shows the measured X-ray structure factors, $|F|^2$, of *n*-doped $SrTiO_3$(001) in air and in electrolyte (both before and after training). The data from samples in air are consistent with previous reports in the literature,[27] and the line-shape is the same for doped and undoped samples. Non-linear least squares fits of $|F|^2$ to atomic models (solid blue line in Figure 2a) suggest that the structure represents a relaxation of the well-known $TiO_2$ double-layer model.[28,29] Immersing the doped substrates in 0.1 M NaOH, however, dramatically alters $|F|^2$ and the resulting (untrained) surface structure is stable in electrolyte under UV illumination. The same structural evolution also occurs in 0.1 M CsOH, suggesting that the counter ion does not play a role in these structural changes. This structural evolution does not occur for undoped samples in either NaOH or 0.1 M $H_2SO_4$. Reports in the literature previously established that photoassisted water splitting at open circuit is negligible for $SrTiO_3$ in $H_2SO_4$, but readily observed in basic medium[14] Figure 2b schematically illustrates the dependence of the structural evolution upon doping and choice of electrolyte.

These experimental studies of the stoichiometry and atomic structure of the untrained surface offer an excellent starting point for interpretation by JDFT calculations. State-of-the-art *ab initio* techniques for study of the electrochemical interface either require computationally prohibitive molecular dynamics to sample the liquid structure, or they consider the surface in vacuum or with only a single frozen layer of water. Our JDFT calculation efficiently includes the atomically-detailed structure of a thermodynamically-sampled liquid by construction.[26] Because JDFT provides full access to the electron density of both the solid surface and the contacting liquid, we calculate $|F|^2$ and compare it directly with experiment. Figure 3b displays the best JDFT candidate structure for untrained $SrTiO_3$ immersed in 0.1 M NaOH, a relaxed 1x1 double-$TiO_2$-terminated structure. The black curve in Figure 3a shows that the *ab initio* $|F|^2$ agrees extremely well with experiment *with no adjustable parameters,* thereby demonstrating the power of the JDFT approach. The JDFT structure describes the X-ray data much more accurately than the 2x1 reconstruction that has the minimum free energy in air or an *ab initio* calculation with a single layer of water molecules adsorbed on the surface (See Figure S13).

As shown in Figure 2a, when an electrode is trained (biased to +0.8 V vs. Ag/AgCl) during UV illumination, the surface evolves to a new stable structure. Trained electrodes exhibit a significant enhancement of $|F|^2$ between the 001 and 002 peaks that is independent of the electrode potential. This enhancement of $|F|^2$ may be characterized by defining a reaction coordinate midway between the two peaks (dotted line in Figure 2a). Figure 2b summarizes the dependence of this enhancement upon doping, choice of electrolyte, and history of applied bias. SECM measurements using a selective $O_2$ microelectrode, under thoroughly de-oxygenated conditions and UV light chopping (Figure 2c), demonstrate a large increase in the photocatalytic activity on trained substrates.[22] SECM can measure the minute changes in $O_2$ evolution with sufficiently high temporal resolution to characterize *in situ* the reactivity of the open circuit $SrTiO_3$ surface just before and after training. Although the activity at open circuit represents only a small fraction of the activity at 0.8V vs. Ag/AgCl, the impact of the subtle surface change on the activity at open circuit is profound. The activity of the trained $SrTiO_3$ surface is $2.6 \pm 0.9$ times larger than that of the untrained surface.[*] (Figure 2d) The same substrates, even after training, were not active towards photocatalytic water splitting in acidic medium, but recovered the enhanced activity if returned to 0.1 M NaOH (Figure 2d and Figure S19). The training-induced structural change survives removal from the electrochemical cell and thorough rinsing with water. Because, at open circuit, transport is not aided by an external field, bulk carrier recombination limits the quantum efficiency. Nevertheless, our result clearly demonstrates the correlation between the structural change in only the topmost atomic layers and the enhanced reactivity.

We explored two possible explanations for the training-induced structural change: (1) the formation of a photogenerated surface intermediate O(H) ad-layer; and, (2) a change in crystal structure of the surface. Pursuing the ad-layer hypothesis, we performed surface interrogation SECM, [23,24] which demonstrated that a limiting coverage, consistent with one equivalent of reactive O (or two OH) per unit cell of $SrTiO_3$, is reached at positive potentials (see supplementary documents). However, non-linear least squares (NLS) fits of the X-ray data to atomic models indicate that the addition of an O(H) ad-layer alone cannot explain the observed change of $|F|^2$, particularly between the 001 and 002 peaks.

---

[*] Training increases the relative quantum efficiency (ratio of the efficiency at open circuit to the efficiency at 0.8V vs. Ag/AgCl) from 0.8% to 2.1%. These numbers are from measurements on five samples and at least two sites on each.

To explore the structure change hypothesis, we relaxed multiple (~100) surface stoichiometries and geometries within JDFT and found that the only likely candidate for the trained surface is an oxygen-deficient, biaxially strained, anatase-like structure, as shown in Figure 3d. Since anatase is also a catalyst for photoassisted water splitting, it is perhaps not surprising that the trained surface is associated with enhanced reactivity. At least one layer of Sr atoms has been removed, leaving a triple $TiO_2$ termination and explaining the observed irreversibility of the training. This candidate for the trained structure has the Ti-stacking of anatase rather than $SrTiO_3$, and it is oxygen deficient at the surface. The fully *ab initio* calculation which agrees best with the X-ray data for the trained surface has the stoichiometry of $Ti_2O_2$ at the surface with a capping hydroxide layer. However, there is still significant discrepancy between the JDFT calculation with no adjustable parameters and the experimental structure factor. (See the dotted line in Figure 3c)

A JDFT-guided nonlinear least squares fit to the experimental structure factor allows the atoms to move away from their minimum energy positions to consider the effect of non-equilibrium or thermal processes, and to develop partial occupancies to account for missing surface atoms and other disorder. Adding a generalized penalty function to the traditional chi-squared fit allows exploration of other surface configurations with energies within approximately *kT* (room temperature) per atom of the JDFT minimum energy. The solid red curve in Figure 3c applies this JDFT-guided fit procedure to the above-described fully *ab initio* structure. This particular fit prevents atoms from moving more than 0.22 Å from their JDFT positions and has an excitation energy of less than 2 *kT* per fit atom. The partial occupancy of Ti atoms in this fit suggests a (still O-deficient) surface stoichiometry between $Ti_2O_2$ and $TiO_2$. The agreement between the JDFT-guided and experimental structure factors is now excellent, providing convincing evidence that the trained $SrTiO_3$ surface must exhibit an anatase-related structure (as in Figure 3d).

The identification of a highly reactive, anatase-related termination for the $SrTiO_3$ surface after training could have significant impact upon the field of photocatalysis. Going forward, establishing the structural and photocatalytic differences between anatase and the anatase-related surface of $SrTiO_3(001)$ may provide insight into ways to engineer improved photocatalysts. Unlike $SrTiO_3$, anatase has the disadvantage of requiring an applied bias to photo-generate both $O_2$ and $H_2$ from $H_2O$. However, anatase has a smaller band gap than $SrTiO_3$, enabling it to capture more of the solar spectrum. The similarity in the active surface of these established photocatalysts suggests a separation in roles between the surface and the bulk of a designer photocatalyst. For example, combining an anatase-related surface (to provide chemical stability and high catalytic activity) with an underlying host that is better matched to the solar spectrum and can impose optimal biaxial strain on the surface ($SrTiO_3$ imposes 3.2% biaxial tensile strain) is a clear strategy for the future.

This work illustrates the critical role that history-dependent surface structure plays in the photo-catalytic water-splitting activity of SrTiO$_3$. Synthesizing information from *in situ* X-ray surface-sensitive techniques, electrochemical and functional characterization methods, and JDFT calculations, we demonstrate that training the SrTiO$_3$ electrode changes the surface structure from a double layer TiO$_2$ termination to an oxygen-deficient biaxially strained anatase-like structure, and concurrently triples the photoactivity. The profound dependence of electrode structure and reactivity upon the training by applied bias has broad implications beyond just photocatalysis. The insights gained in this work are highly relevant to the design of surface chemical modifications for applications such as pollutant remediation and functional coating, where surface reactivity under zero applied bias is key. Furthermore, the synergistic combination of *in situ* measurement techniques with theory opens a promising path towards fundamental mechanistic understanding of surface reactivity in electrolyte media.

**Acknowledgments**
We thank Jacob Ruff, Darren Dale, and Hanjong Paik for their technical support. This material is based upon work supported as part of the Energy Materials Center at Cornell (EMC2), an Energy Frontier Research Center funded by the U.S. Department of Energy, Office of Science, Office of Basic Energy Sciences under Award Number DE-SC0001086. This work is based upon research conducted in part at the Cornell High Energy Synchrotron Source (CHESS), which is supported, by the National Science Foundation and the National Institutes of Health/National Institute of General Medical Sciences under NSF award DMR-0936384. J.R.L acknowledges the University of Illinois at Urbana-Champaign for start-up funds.

**Author Contributions**
Sample preparation: MP, XH, JRL and DGS
X-ray cell/detector/software development: MP, XH, NLR, JRL, PJYK, HDA, and JDB
Data collection at CHESS: MP, XH, PJYK, JRL, MS, BHS, NLR, HDA, and JDB
SECM measurements: JRL, MP, MS, BHS, NLR, and HDA
JDFT studies: KLW, DG, and TAA
Manuscript preparation: MP, XH, JRL, KLW, NLR, HDA, TAA, DGS, and JDB

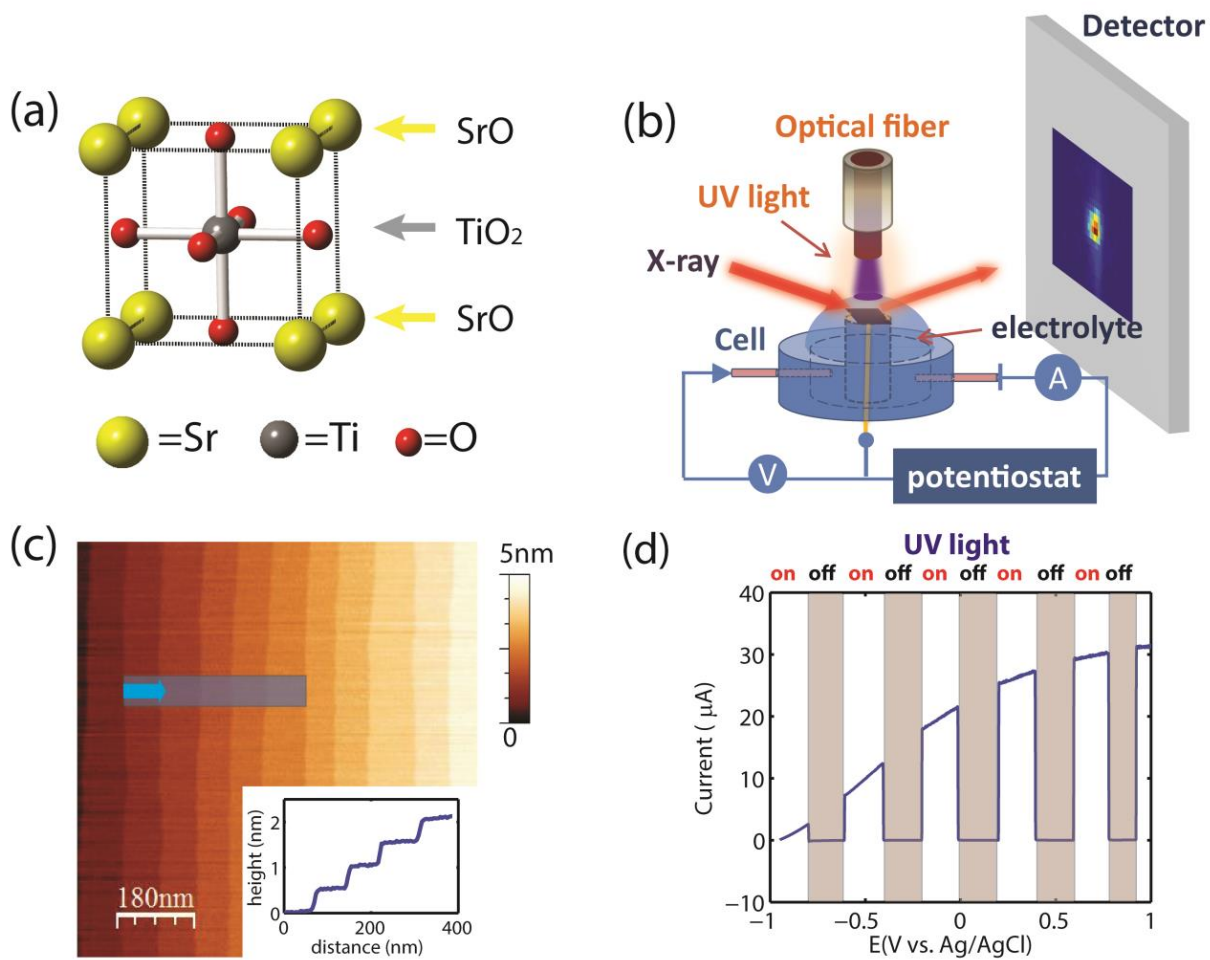

**Figure 1.** **(a)** SrTiO$_3$ unit cell. Crystal structure alternates SrO and TiO$_2$ layers along [001] **(b)** Schematic of the experimental set-up for *in situ* X-ray reflectivity of SrTiO$_3$ during photoassisted (electro)chemistry. **(c)** Tapping mode AFM image of SrTiO$_3$ after surface preparation. The inset shows the height profile of the shadowed area of the AFM image. Atomic terraces are about 0.5 nm in height and 90 nm in width. **(d)** Light-chopped linear sweep voltammogram of SrTiO$_3$ in 0.1 M NaOH illuminated with a 200 W Hg/Xe lamp; electrode area was 0.7 cm$^2$; scan rate is 20 mV/s.

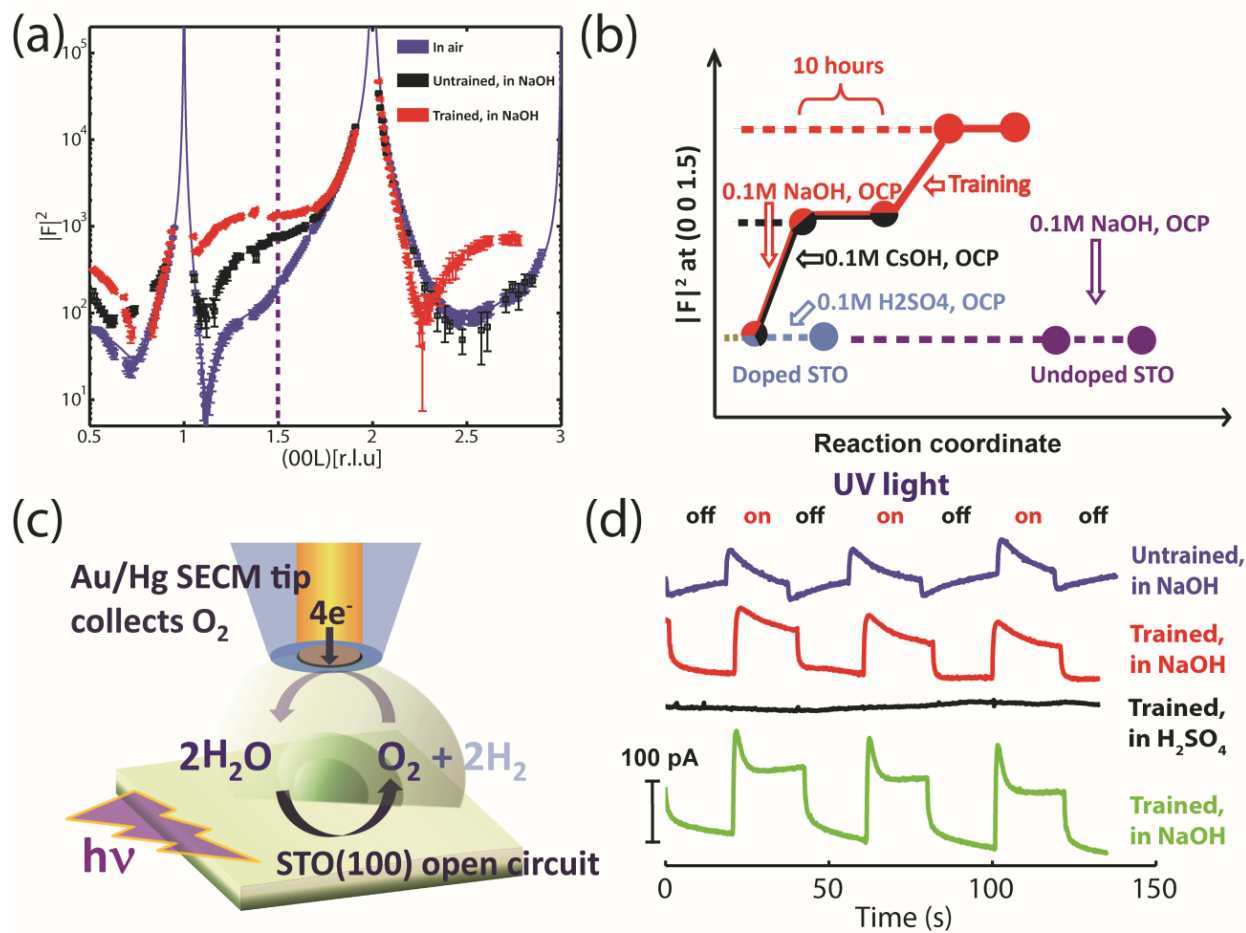

**Figure 2. (a)** 00L structure factor of SrTiO$_3$ in air (blue), in 0.1 M NaOH at open circuit before (black) and after (red) training. Blue solid line is the best fit to the accepted vacuum termination. **(b)** Map to the evolution of $|F|^2$ at (0 0 1.5) r.l.u. of samples in different electrolytes: Red, doped SrTiO$_3$ in 0.1 M NaOH; Black, doped SrTiO$_3$ in 0.1 M CsOH; Blue, doped SrTiO$_3$ in 0.1 M H$_2$SO$_4$; Purple, undoped SrTiO$_3$ in 0.1 M NaOH. **(c)** SECM in O$_2$ substrate collection mode. Hg/Au amalgam tip detects oxygen produced by the water splitting reaction at the SrTiO$_3$ electrode at open circuit. **(d)** SECM collection with UV light on/off: (blue) 0.1 M NaOH before training, (red) after training (biasing to 0.8 V vs. Ag/AgCl for 40 min). (black) upon immersion in 0.1 M H$_2$SO$_4$ and (green) after returning to 0.1 M NaOH. The oxygen generation rate is proportional to the current, 100 pA ~ 200 µmol h$^{-1}$m$^{-2}$ (see Figure S19).

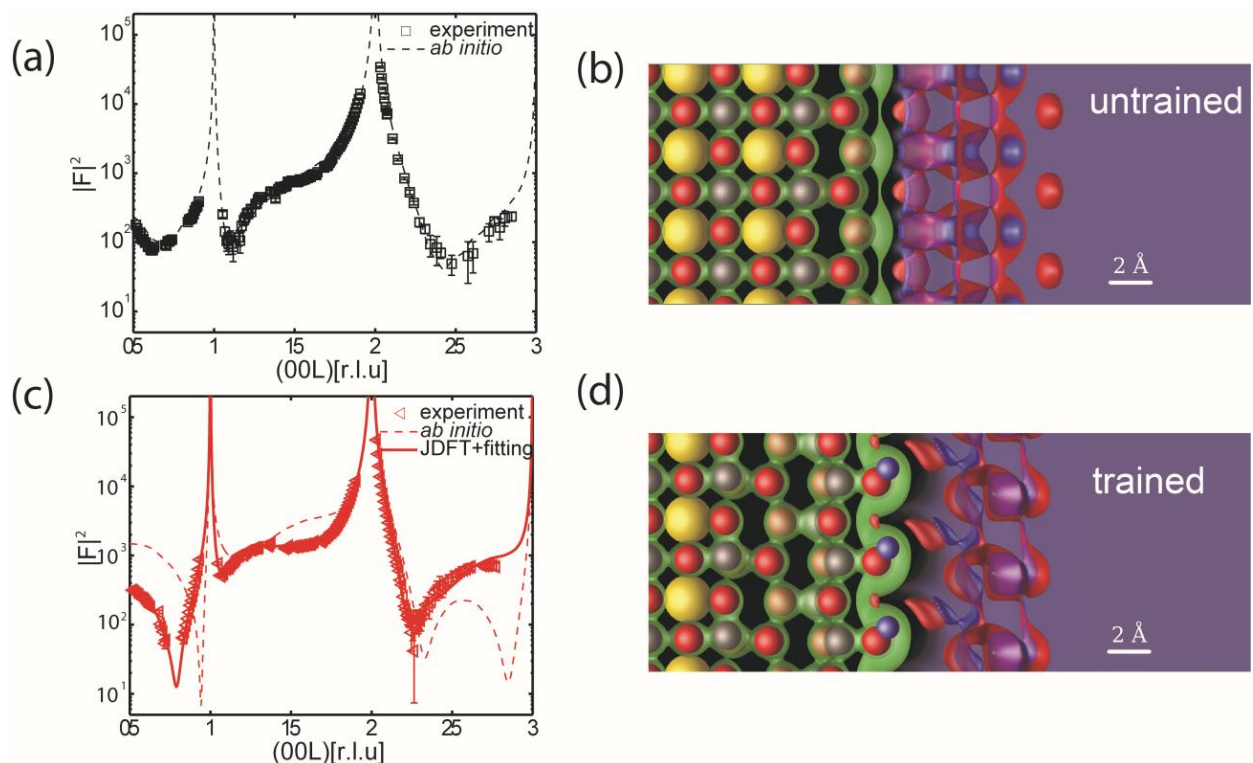

**Figure 3. (a)** 00L structure factor of SrTiO$_3$ in 0.1 M NaOH, untrained. Black square is experimental data. Black dash line is the structure factor calculated from the JDFT structure, with no adjustable parameters. **(b)** The JDFT structure for the untrained surface of SrTiO$_3$. **(c)** 00L structure factor of SrTiO$_3$ in 0.1 M NaOH, trained. Red triangle is experimental data. Red dash line is from JDFT structure. Red solid line is a fit constrained to the JDFT structure with a penalty function. **(d)** The fully *ab initio* JDFT structure for the trained surface of SrTiO$_3$. Yellow spheres are strontium, red are oxygen, blue are hydrogen, and silver are titanium atoms. Green, red, and blue density contours represent electron, oxygen, and hydrogen density respectively as described in supplementary documents.

# Supplementary Information for

**Structure of the photo-catalytically active surface of SrTiO$_3$**

Manuel Plaza,[1,µ] Xin Huang,[µ] J. Y. Peter Ko,[2] Joel D. Brock[*]
[2]School of Applied and Engineering Physics, Cornell University, Ithaca, NY 14853, USA

Mei Shen, Burton H. Simpson, Joaquín Rodríguez-López[µ]
Department of Chemistry, University of Illinois at Urbana-Champaign, Urbana, IL 61801

Nicole L. Ritzert, Héctor D. Abruña
Department of Chemistry and Chemical Biology, Cornell University, Ithaca, NY 14853, USA

Kendra Letchworth-Weaver,[µ] Deniz Gunceler, and T. A. Arias
Department of Physics, Cornell University, Ithaca, NY 14843, USA

Darrell G. Schlom,
Department of Materials Science and Engineering, Cornell University, Ithaca, NY 14853, USA

[1]Current Address: Departamento de Física de la Materia Condensada, Universidad Autónoma de Madrid, Ciudad Universitaria de Cantoblanco, 28049, Madrid, Spain
[2]Current Address: Cornell High Energy Synchrotron Source, Cornell University, Ithaca, NY 14853, USA
[µ]Equal contribution
[*]Corresponding Author

Correspondence to: jdb20@cornell.edu

**This PDF file includes:**

# Methodology for crystal truncation rod X-ray measurements

For a perfect crystal with infinite size, reciprocal space features periodic, delta-function Bragg points. However, when the crystal presents an atomically smooth surface, the Bragg points are extended along the surface normal direction, becoming rods. These rods are called Crystal Truncation Rods (CTRs)[1,2]. The specular CTR corresponds to the x-ray reflectivity (XRR). The intensity distribution along a CTR is given by the structure factor, ($|F|^2$), and is intimately related to the real space structure of the surface.

Here, *in situ* Surface X-Ray Diffraction (SXRD) experiments were conducted at the Cornell High Energy Synchrotron Source (CHESS) A2 station. The X-ray energy was 30 keV. X-ray intensities were collected by a 41cm x 41cm General Electric (GE) area detector (DXR250RT, GE Inspection Technologies). Each pixel is 200 micron x 200 micron. The distance between the sample and the detector is 927mm. Each exposure takes 50 seconds. Figure S1(a) shows how several CTRs are measured simultaneously using a large area detector. The (00L) (specular) and (10L) (off-specular) CTRs are shown in the diagram. The intersection of a CTR with the Ewald sphere produces a small spot on the detector. Small regions of interest on the detector containing the CTR signals are selected: $ROI_1$ (for (00L) CTR) and $ROI_2$ (for (10L) CTR), as an example. Figure S1(b) shows one of the $ROI_1$ images, containing the (00L) CTR spot when L=1.75. The integrated intensity of the CTR spot is obtained after subtracting the diffuse scattering background. After correcting for attenuation in the solution and the silica dome of the electrochemical cell, the structure factor is obtained. More details about the calculation can be found in Reference [3]. The (00L) CTR of a fresh *n*-doped $SrTiO_3$ (001) (STO) crystal in air resulting from this procedure is shown in Figure S1 (c) as an example.

# Sample preparation and photoelectrochemical characterization

*Chemicals*

All chemicals for electrochemical measurements were used as received. Sodium hydroxide (NaOH, semiconductor grade, 99.99%) was from Aldrich Chemical Co. (Milwaukee, WI). Potassium hexacyanoferrate(III) (ferricyanide, $[Fe(CN)_6]^{3-}$) (ACS grade) was obtained from Fisher Scientific (Fairlawn, NJ). Mallinckrodt-Baker (Phillipsburg, NJ) was the source of potassium nitrate. Water (18 MΩ•cm) from a Millipore (Billerica, MA) Milli-Q system was used to clean all glassware and to prepare all aqueous solutions. Cesium hydroxide (CsOH) solutions were prepared from a 50% wt. solution in water (99.9%, Sigma-Aldrich, St. Louis, MO). Sulfuric acid ($H_2SO_4$) solutions were prepared from a 99.9999% solution (92% min, Alfa Aesar, Ward Hill, MA). Strontium titanate (STO) samples were purchased from MTI Corp. Acetone (from Macron Chemicals), Isopropyl Alcohol (from Mallinckrodt Chemicals) and 7:1 $NH_4F$-HF buffered oxide etchant solution (from J. T. Baker, VLSI grade) was used in STO surface etching process.

*Strontium titanate electrodes*

As-received STO(100) samples were first cleaned with acetone and isopropyl alcohol and then sonicated in deionized water for 15 min. After cleaning, the samples were etched with 7:1 $NH_4F$-HF buffered oxide etching (BOE) solvent for 30s, and then rinsed with abundant deionized water. Finally, the samples were annealed at 1000 °C for 2.5 hr in oxygen atmosphere, and then slowly cooled to room temperature. The next step in the preparation was generating oxygen vacancies by annealing the samples at $10^{-7}$ Torr at 1000 °C for 1 hr. UV-Vis measurements (shown in Figure S2) were performed using a Shimadzu UV3600 spectrophotometer in reflection configuration. An undoped STO sample was used as reference for the differential measurements. Samples absorbed light well into the UV region of the spectrum and appeared dark to the eye.

The doping level of the samples was estimated by Hall effect measurements using a cryogenic superconducting quantum interference device (SQUID) magnetometer MPMS-XL by Quantum Design in a 4 K He environment. Indium contacts to the sample ensured ohmic contacts. The concentration of the oxygen vacancies was $4 \times 10^{17}$ cm$^{-3}$. Electrochemical impedance analysis confirmed this doping level. The electrochemical measurements were performed using doped STO samples as the working electrode together with a Ag/AgCl in saturated KCl reference electrode bridged with a 0.2 M KNO$_3$ + 3% agar salt bridge (to prevent chloride contamination in the cell) and either a large area Au or Pt counter electrode. Electrochemical impedance analysis using a Model SI 1280B Electrochemical Measurement Unit (Solartron Analytical, Farnborough, Hampshire, UK) via Mott-Schottky analysis was performed in dark conditions in 0.1 M NaOH. Typical Mott-Schottky plots of STO samples annealed at 1000 °C are shown in Figure S3. Under majority carrier depletion conditions, the imaginary component of the cell impedance can be assumed to represent the space charge capacitance[4]. The slope of the Mott-Schottky plot indicates the type and density of majority carriers, as expressed in Equation S1 (at 298 K, with $n_D$ in cm$^{-3}$, $E$ in V and $C_{SC}$ in µF cm$^{-2}$):

$$\frac{1}{C_{sc}^2} = \left[\frac{1.41 \times 10^{20}}{\varepsilon n_D}\right][E - E_{fb} - 0.02571] \tag{S1}$$

Here $E_{fb}$ stands for the flatband potential of the sample and $\varepsilon = 332$ is the relative permittivity of STO[5]. These plots show that the samples obeyed the conditions for the validity of Equation S1, where the capacitive response is independent of frequency in a suitable range and where a linear response is observed between the capacitive function and the applied potential[4]. This linear dependence is observed at potentials positive of $E_{fb}$ and therefore corresponds to n-doping. The doping levels are consistent with those measured by Hall Effect, where $n_D \sim 10^{17}$ cm$^{-3}$. While this doping level is smaller than that used by other preparation techniques, such as annealing in hydrogen where typical $n_D \sim 10^{19}$ cm$^{-3}$ [5], the electrochemical response as judged from CV experiments (main text) and Mott-Schottky plots is adequate for our *in situ* X-ray and SECM experiments.

*Photoelectrochemical measurements*

The photoelectrochemical response of the STO (100) samples was characterized and controlled using a monochromated UV light source. We used a model 6291 200 W Hg(Xe) Lamp (Oriel Instruments, Stratford, CT) in a Spectra-Physics (Santa Clara, CA) Model 66902 Lamp Housing powered by a Spectra-Physics Model 69911 Power Supply operated at 200 W. A two-meter long Oriel Model 78278 VIS-NIR Single Fiber Cable (core diameter of 1000 µm), connected using a Model 77776 Fiber Bundle Focusing Assembly, was used to irradiate the top of the sample through the electrolyte solution. An Oriel Model 7240 monochromator with a resolution of ±20 nm (±0.1eV approx.) was used to select the wavelength.

A first test to confirm the photoelectrochemical response of the STO samples was to measure their open circuit potential. Figure S4 shows a typical open circuit transient upon illumina-tion with the full spectrum of the UV-Vis lamp. Typical dark open circuit potentials were between -0.7 and -0.6 V vs. Ag/AgCl and responded immediately to light excitation to values between -1.0 and -0.9 V vs. Ag/AgCl. The use of a 400 nm cutoff filter demonstrates that the samples responded almost exclusively in the ultraviolet portion of the spectrum, Figure S5.

Photoaction spectra, Figure S6, were obtained in 0.1 M NaOH electrolyte under illumination with the Hg(Xe) lamp and at a potential of 0.8 V vs. Ag/AgCl to characterize the conditions utilized during synchrotron experiments. The observed photocurrent maximum at λ = 390 nm (~3.2 eV), likely the product of a spike in the Hg(Xe) lamp output and the indirect bandgap transition in $SrTiO_3$, was used in our following measurements to ensure maximum activity under controlled illumination conditions. A change in approximately four orders of magnitude in current density between the dark condition and under illumination indicate the high activity of our samples and their adequacy for further structural characterization during electrochemical activity.

In order to determine the absolute photocatalytic efficiency of our system, we measured the Incident Photo Conversion Efficiency (IPCE) corrected for reflective losses (Figure S2) for a representative strontium titanate sample with a doping density in the $10^{17} cm^{-3}$ range. IPCE measurements were performed using a 300 W Xe lamp (6258 Oriel) with nearly constant irradiance in the 300 nm to 800 nm range. We determined the light output at each wavelength using a calibrated photodiode [Melles Griot 13DAH005] with a typical response of 0.3 A/W. Figure S7 shows the IPCE obtained for a typical $SrTiO_3$ sample, and it is evident that the IPCE drops off at higher light energy. The photodynamics of oxygen vacancy $SrTiO_3$ samples have shown increased recombination rates when the photon energy increases beyond the energy required to excite the material[6]. For instance, the photoluminescence decay times for strontium titanate samples with oxygen-vacancies (obtained through Ar-irradiation) with the same doping density as ours, decrease 40-fold in magnitude when going from illumination energies of 3.44 eV to 3.87 eV (360 nm to 320 nm). This is consistent with our IPCE results, nonetheless, these effects do not to weaken our observation that our samples are highly active towards photocatalysis under UV illumination.

## *In situ* CTR measurements

Doping with oxygen vacancies is necessary to increase the electrical conductivity of the STO. CTRs of STO before and after the oxygen vacancy generation process ($n_d$ =$4x10^{17} cm^{-3}$) are shown in Figure s8. The CTRs are indistinguishable, demonstrating that the doping process does not alter the structure of the (001) STO surface.

After immersion in 0.1 M NaOH and resting at open circuit under UV illumination, the surface evolves to a new structure as demonstrated by the change of the CTR line shape, Figure 2(a) in the main text. This new structure is stable for many hours. To demonstrate this stability, we measured the (00L) CTR four times consecutively, as shown in Figure s9(a). The line shape of the CTR did not change significantly during the measurements (10 hours in total).

The sample was then biased to +0.8V vs. Ag/AgCl. The surface evolved once again to another new surface, different from the two previous ones. Figure s9(b) shows the (00L) CTR of a STO sample under various conditions: (1) biased to +0.8V vs. Ag/AgCl; (2) after returning to open circuit potential (OCP); and, (3) at OCP 3 hours later. The CTR profile is indistinguishable in all three cases, indicating that this final structure is stable and does not return to the original OCP structure.

Undoped STO behaves differently. No CTR evolution occurs on undoped STO when it is immersed in basic electrolyte, as demonstrated by the CTRs shown in Figure s10(a).

One possible explanation for the surface evolution is irreversible cation adsorption on the STO surface. To test this possibility, we repeated the CTR measurements in basic solutions using 0.1 M CsOH, which has a different cation size but similar chemical properties to NaOH. Figure s10(b) compares the (00L) CTR of STO in air and in 0.1 M CsOH at OCP. A similar CTR profile change to that in 0.1 M NaOH occurs in 0.1 M CsOH, indicating that the surface evolution is not a result of cation adsorption processes.

As a comparison to basic electrolytes, we performed experiments in acid electrolyte. The STO surface is not active towards photocatalytic water splitting at open circuit in acid solution (see Figure 2(d) in the main text). CTR measurements were conducted by immersing a freshly prepared STO sample in 0.1 M $H_2SO_4$ at OCP, and then comparing the CTR measurements before and after immersion (Figure s10(c)). The CTR profiles are quite similar before and after immersion in the acid electrolyte solution, indicating that the acid medium does not change the STO surface.

In summary, oxygen vacancies, UV illumination and basic conditions are required to reconstruct the surface to the catalytically active state.

## CTR interpretation

In order to obtain the real space structure from the experimental $|F|^2$, the data were fit to several atomic models by non–linear least square fitting. The atomic positions, Debye-Waller factors, and site occupancy were all relaxed. The reduced chi squared ($\chi^2$) was used as the goodness of fit parameter: $\chi^2 = \frac{1}{N-n-1}\sum_{i=1}^{N}\frac{(y_i - f(x_i))^2}{\sigma_i^2}$. Here, N is the total number of data points; $n$ is the number of adjustable parameters; $\sigma_i$ is the experimental error at $x_i$, mainly from the stochastic noise; $y_i$ is the measured experimental data at $x_i$; and, $f(x_i)$ is the calculated value at $x_i$. The fits provide a good suggestion for the surface structure. However, because we vary both the atomic species and the site occupancy, the atomic model is not unique. We use Joint Density Functional Theory calculations to ensure that the fitted surface structure is physically realizable.

Figure 2(a) in the main text shows both experimental (00L) CTR's under three different conditions and the best fit to the CTR of the surface in air (blue). The CTR prediction for the untrained surface freshly immersed in aqueous electrolyte (black curve in Figure 3(a)) has no fitted parameters whatsoever. Figure 3(c) shows the best fit to the CTR for the trained surface in aqueous electrolyte after biasing (red). The fit to the CTR in air minimizes only $\chi^2$ while the fit to the trained surface CTR minimizes both $\chi^2$ and a penalty function which prevents unphysical variation from the JDFT structure (as described below). The fits are in excellent agreement with the experimental data. The $\chi^2$ values range from 1 for the surface in air to 7 for the JDFT-guided fit to the trained surface.

## Joint Density Functional Theory

Joint density-functional theory[7] states that the free energy of any quantum-mechanical system in contact with a liquid environment is given by

$$G([n(r),\{N_\alpha(r)\}] = A_{KS}[n(r)] + \Omega_{lq}[\{N_\alpha(r)\}] + \Delta A[n(r),\{N_\alpha(r)\}] \quad (S2)$$

where $n(r)$ is the local ground state electron density of the quantum-mechanical system, $\{N_\alpha(r)\}$ are the ground state densities of each of the independent atoms in the fluid (α=O,H for water), and each of the three terms is a functional which must be approximated. The term $A_{KS}[n(r)]$ is the standard Kohn-Sham functional for a quantum-mechanical system in vacuum, in this work described by the generalized gradient approximation.[8] The liquid environment is described by the classical DFT functional $\Omega_{lq}[\{N_\alpha(r)\}]$, which provides by construction all thermodynamic information included in the equation of state for a particular liquid.[9,10] The interaction between the liquid environment and the quantum-mechanical system is included within the coupling functional $\Delta A[n(r), \{N_\alpha(r)\}]$ approximated by a density-only DFT.[11] Variational minimization of the above free energy functional leads to an atomically detailed description of the liquid environment at the surface while avoiding the empiricism and thermodynamic sampling required by the alternative approaches. The JDFT framework, even with a simplified description of the liquid environment (see below), has been proven reliable for mapping electrochemical observables to *ab initio* computables.[12] The above theoretical method is implemented efficiently with hybrid MPI and gpu parellelization in the open source code JDFTx.[13]

Most other solvation theories that avoid thermodynamic sampling model the liquid environment as a simple continuum dielectric, which forms a cavity around the solute system (in this case the SrTiO$_3$ electrode surface). The only fluid description available in these theories is a simple electrostatic bound charge, the simplest representation of the true shell structure present in any realistic fluid. These continuum solvation theories may be further divided into traditional Polarizable Continuum Models (PCM's)[14], which use multiple solute-specific parameters to determine the cavity around the solute, and *implicit* or iso-density continuum solvation theories,[15,16] which describe the fluid based on a single variable (the quantum-mechanical electron density $n(r)$ of the electrode surface). *Implicit* solvation theories may be placed in the rigorous, and in-principle exact JDFT framework,[7] and used successfully for electrochemistry;[12,17] however, significant approximations are made in any practical *implicit* JDFT calculations.

In contrast, the theory used for the first time in this manuscript is *explicit* JDFT, which contains full, atomically detailed information about the structure of the liquid at the surface. There are independent variables describing both the surface and the fluid in the theoretical calculation. The surface is described by its quantum-mechanical electron density $n(r)$ in the traditional Kohn-Sham functional $A_{KS}[n(r)]$. The fluid molecules are described by one scalar field $\mu(r)$ (local chemical potential for a water molecule) and one vector field $\vec{\epsilon}(r)$ (local electric field experienced by a water molecule).[10] The coupling interaction between the fluid and the surface in $\Delta A[n(r), \{N_\alpha(r)\}]$ is computed including both electrostatic and quantum-mechanical effects. The interactions among the fluid molecules included in $\Omega_{lq}[\{N_\alpha(r)\}]$ include electrostatic effects, hard sphere packing, and additional terms constrained to reproduce the bulk equation of state, liquid-vapor coexistence, and surface tension of real liquid water. Thus, unlike any *implicit* solvation theory, *explicit* JDFT can predict the solvation shell structure of the liquid from the atomic densities $\{N_\alpha(r)\}$ where α = O, H for water.[9]

*JDFT calculations of X-ray signatures*

JDFT captures the atomic details of the thermodynamically averaged interfacial liquid structure and provides highly accurate predictions of solvated atomic structure and the electronic properties of solvated systems. Once the relaxed atomic positions, electron density, and fluid density profiles are determined from a JDFT calculation with a given stoichiometry and surface configuration, we may use them to compute the X-ray structure factor $F(\vec{q})$, with no adjustable parameters. We do not require tabulated form factors, nor is there a need to guess the relevant oxidation states of the surface atoms. The main quantity required for a completely first principles prediction of a CTR is the total electron density (including valence, core, and fluid electrons). We calculate $n_{tot}(\vec{r})$ from JDFT using Equation S2 and average it over the plane parallel to the surface to find $\bar{n}_{tot}(z) = \iint n_{tot}(\vec{r}) dx dy$. With this electron density profile, we compute the structure factor for a specular X-ray CTR ($\vec{q} = q_\perp = [0\ 0\ 2\pi L/a_0]$) as

$$F(\vec{q}) = \frac{\int_0^{a_0} \bar{n}_{bulk}(z) e^{iq_\perp z} dz}{1 - e^{iq_\perp a_0}} + \int_0^{a_z} \bar{n}_{tot}(z) e^{iq_\perp z} dz + \frac{n_{fl}^{(b)} e^{iq_\perp a_z}}{iq_\perp} \quad (S3)$$

Figure s11 shows a schematic representation of this calculation. The first term in Equation S3 represents the ideal truncation (by a geometric series) of the perfect crystal, which is assumed to extend from $-\infty < z < 0$. We perform a calculation of the bulk crystal with DFT lattice constant $a_0$ to calculate the planar average density $\bar{n}_{bulk}(z)$. The final term, which assumes bulk fluid with constant electron density $n_{fl}^{(b)}$ from $a_z < z < \infty$, is present to prevent edge effects from the fluid, similar in function to the first term. The second term, which includes all the important interfacial effects, is simply the Fourier transform of $\bar{n}_{tot}(z)$ from the JDFT calculation of the surface. The total electron density includes several key components: 1) the valence electron density $n(\vec{r})$ of the quantum mechanical electrode, 2) the valence electron density $n_{fl}(\vec{r})$ of the fluid, determined by the atomic fluid density fields and a single quantum mechanical calculation for each type of fluid molecule, and 3) the core electron density $n_{core}(\vec{r})$, determined by the all electron calculations used to generate the atomic pseudopotentials.

*First Principles CTR Predictions for SrTiO$_3$*

With the combination of the JDFT framework in Equation S2 and the recipe for calculating crystal truncation rods from first principles in Equation S3, we are able to make meaningful theoretical predictions of *operando* X-ray signatures. JDFT provides a prediction of not only the interfacial liquid structure, but also the effect of the screening from the liquid environment on the electronic structure of the electrode and the positions of each atomic layer. JDFT can also predict hydrogen bonding and physisorption effects quite accurately, so associative water adsorption is captured within the classical liquid functional. However, covalent bonds or charge transfer reactions between the liquid molecules and electrode must be included in the quantum mechanical portion of the calculation. For each surface termination or stoichiometry considered, we must thus consider hydroxide, water, and oxygen chemisorption quantum-mechanically. Additionally, reconstructions parallel to the surface which result from rearrangements of the electrode atoms must be considered explicitly (though the classical DFT fluid accounts for partial coverages of water, on even a single unit cell of SrTiO$_3$). All these degrees of freedom make this endeavor quite daunting, and virtually impossible to approach with molecular dynamics due to the addi-tional complication of thermodynamically sampling the liquid. JDFT is an ideal technique for this problem because it is both computationally efficient and provides atomic details.

To consider all potential $SrTiO_3$ surface terminations which could be participating in the water splitting reaction, we performed a large scale JDFT search over many stoichiometries and surface compositions.[18] We tried surface structures with different number of terminating SrO or $TiO_2$ layers and horizontal reconstructions (1x1, 2x1 and $\sqrt{2}$x$\sqrt{2}$), including several structures presented in the literature.[19-21] Because covalent bonds or charge transfer reactions between the liquid molecules and electrode must be included in the quantum mechanical portion of the calculation, each surface was considered with adsorbed species involved in the water-splitting reaction ($O^{2-}$, $OH^-$ and $H_2O$). We choose not to include the sodium cation within the quantum-mechanical portion of the calculation because the experimental measurements in Figure s10(b) show that substitution of cesium for sodium does not change the X-ray signature of the untrained surface. Because the high atomic number of cesium should significantly alter the structure factor if the cation participated in the interface structure, we surmise that the influence of the cation is unimportant. The JDFT calculations were performed at varying levels of approximation for the liquid; those surfaces which appeared to be good candidates for the CTR from a simple fluid model[17] were then calculated in the more advanced models.[10,13] Those surfaces which were local (but not global) free energy minima were considered, since the water splitting reaction is a non-equilibrium process with multiple intermediate states.

For the untrained $SrTiO_3$ surface freshly immersed in a liquid electrolyte, JDFT-calculated crystal truncation rods (CTR's) agree with X-Ray reflectivity measurements with no parameters whatsoever fit to the experimental X-Ray data. The surface configuration which agrees best (CTR shown in Figure 3(a) and structure shown in Figures 3(b) and Figure s12(a)) is a double $TiO_2$ terminated surface with a single unit cell parallel to the surface, and without specific chemical adsorption (only physisorbed water). The *z*-positions in lattice coordinates for each atom in this structure are tabulated in Table S1. This surface possesses a different symmetry and is quite distinct from the 2x1 reconstruction which is a free energy minimum in vacuum or air calculations.[19,22] Despite its simplicity, this surface composition matches the experimental X-ray data much more closely than any of the more complicated reconstructions considered. Because the CTR's are highly dependent upon the slight displacements of atomic layers away from their bulk positions, the excellent agreement with the experiment indicates the precision of the JDFT-predicted atomic positions and liquid structure.

Figure s13(a) indicates the superior agreement for the CTR prediction when including the thermodynamically sampled liquid within JDFT, compared to including a single explicit layer of water in vacuum. The vacuum calculation shown by the dotted green line in Figure s13(a) includes relaxed water molecules which dissociate, covalently bond with the surface, and alter the atomic structure, while the JDFT calculation assumes liquid water with shell structure. For reference, Figure s13(a) also shows the vacuum DFT predicted CTR for the 2x1 surface reconstruction[19] which has been shown to possess the minimum free energy in vacuum. Clearly, traditional vacuum DFT techniques are unable to capture the structure of the $SrTiO_3$/electrolyte interface before a bias has been applied.

It was a greater challenge to identify the structure of the more complex trained $SrTiO_3$ electrode surface, which reorders after biasing under water-splitting conditions. Of the approximately 100 surface calculations performed, most failed to agree with the experimental CTR for the activated surface, with the difference being most striking in the middle region, between the first and second Bragg peaks (measured by the "reaction coordinate" $|F|^2$ at $\vec{q} = \frac{2\pi}{a_0}[0\ 0\ 1.5]$). Finally, a structure with a triple layer Ti-rich termination yielded a promising reaction coordinate. This surface is anatase-like in the stacking of the Ti atoms, but is under a strain of over 3% from the anatase lattice parameter and is non-stoichiometric. The best-fitting structure consists of two layers of $TiO_2$, then a layer of $Ti_2O_2$, with one hydroxide bonded to the outermost titanium as shown in Figures 3(d) and s12(b). The predicted CTR for this structure with no fit parameters is shown by the red, dotted line in Figure 3(c). While this prediction is clearly not an exact match to the experimental data, our exhaustive search for alternatives (as well as the success of the JDFT-guided fitting procedure described below) leaves us confident that the discrepancies are due to non-equilibrium processes, defects, or disorder at the surface.

*JDFT-guided Nonlinear Least-Squares Fits*

Nonlinear least-squares fitting allows the atoms to move away from their JDFT minimum energy positions to consider the effect of non-equilibrium processes and the resulting partial occupancies and Debye-Waller factors account for defects and disorder. We may ensure the physicality of these fits by minimizing a residual $R^2$ which includes both $\chi^2$ and a penalty function to prevent the fit positions $\{\zeta_I\}$ from varying significantly from the JDFT predicted positions $\{Z_I\}$

$$R^2 = \chi^2 + \frac{1}{2}\gamma \sum_I (\zeta_I - Z_I)^2.$$

The constant $\gamma = 10\ \text{Å}^{-2}$ determines the relative weight of the penalty function compared to $\chi^2$. The structure factor for this fit is calculated from tabulated form factors for the explicit atoms located at fit positions $\{\zeta_I\}$ and the electron density of the fluid predicted from JDFT. For the activated surface, the best fit positions, occupancies, and Debye-Waller factors, as well as the discrepancies from JDFT-determined positions, are tabulated in Table S2. One significant feature of this fit is the partial occupancy of the top Ti atom in the structure – Ti is only present in half the unit cells, leaving a surface structure which is still slightly oxygen deficient. The corresponding CTR prediction for the best fit structure of the trained surface is shown by the red, solid lines in Figure 3(c). The best fit structure has a total DFT-calculated distortion energy away from the *ab initio* minimum energy structure of only 0.73 eV, corresponding to 0.045 eV (less than 2 $kT$ where $T$ is room temperature) per atom included in the fit. The distortion thus is within the uncertainties of the approximate functionals used in this work. The rms error in the best fit positions compared to the JDFT-predicted positions is

$$z_{rms} = \sqrt{\frac{1}{N_I}\sum_I (\zeta_I - Z_I)^2} = 0.11\ \text{Å}$$

with a maximum change of 0.22 Å for a single atom.

There is also some resulting uncertainty in the fluid structure due to flexibility in our construction of the functional $\Delta A$, which couples the surface and the fluid. In the spirit of ensemble error analysis,[23,24] Figure s13(b) indicates the corresponding range of JDFT predictions for the CTR's in Figures 3(a) and (c). The qualitative agreement with experimental data remains good for all plausible choices of JDFT coupling functional.[23,24]

*Numerical and Computational Details*

To perform a full JDFT calculation of the surface in contact with a liquid environment, we must specify the liquid functional $\Omega_{lq}[\{N_\alpha(r)\}]$ and the coupling functional $\Delta A[n(r), \{N_\alpha(r)\}]$ in Equation S2. For $\Omega_{lq}$, we used a classical liquid functional for water[10] with only rotational contributions to the dielectric response. (Polarizability contributions to dielectric response are negligible for water and slow convergence significantly due to 3 extra degrees of freedom.) We sampled the orientation probability of the water molecule in an ideal gas representation[10] and used 144 quadrature points for the final density profiles shown in Figures 3(b) and (d). For $\Delta A$, to couple the fluid to the quantum-mechanical SrTiO$_3$ electrode, we use 1) the mean-field coulomb interaction between the charge density of the surface and the charge density of the fluid[9] and 2) density-only DFT between the valence electrons of the fluid $n_{fl}(\vec{r})$ and the valence electrons of the surface $n(\vec{r})$. We use the Thomas-Fermi[11] orbital-free expression for the kinetic energy of the electrons and the local density approximation[25] for the exchange and correlation.

All numerical calculations were performed within the DFT++ framework[26] as implemented in the open-source code JDFTx[13] using direct minimization via the conjugate gradients algorithm.[27] We employed the generalized-gradient approximation[8] using a plane-wave basis and a planewave energy cutoff of 800 eV. We used fully periodic boundary conditions and a single unit cell of SrTiO$_3$ for the bulk calculation with a 4x4x4 kpoint mesh[28] to sample the Brillouin zone. The DFT calculated lattice parameter of SrTiO$_3$ was determined from the bulk calculation to be $a_0 = 3.94$ Å, less than 1% above the experimental lattice constant. All calculations presented employ optimized norm-conserving Kleinman-Bylander pseudopotentials.[29,30] A partial core correction[31] was required for the Sr and Ti pseudopotentials. Energy convergence with planewave cutoff and k-point sampling was carefully checked using the bulk system as a test case.

The supercells for the SrTiO$_3$ surfaces were constructed as demonstrated in Figure s12 for the (a) untrained and (b) trained surfaces. We used the minimal real-space unit cell in the $x, y$ directions (parallel to the surface) but maintained k-point sampling commensurate with the bulk calculation (A 1x1 real-space cell with 4x4x1 k-point sampling for the examples shown). All unit cells were constructed to be inversion symmetric about $z = 0$ with a distance of $2a_z \approx 40a_0$ between periodic images of the surface. Note that such a large distance was employed to allow for structuring in the fluid densities at the surface. Some of the high-throughput screening calculations in vacuum and in simplified fluid theories without shell structure[17] instead use coulomb truncation to prevent image interaction. We obtain the geometry of the SrTiO$_3$ surfaces directly from the relaxed nuclear positions $\{R_I\}$ of the JDFT calculation. The total forces on the atoms were relaxed to less than 5 meV/Å.

*Real-space Image Preparation*

The real-space images in Figures 3(b) and (d) are each $3a_0$ by 30 Å. We include the Sr, Ti, O, and H atoms of the quantum mechanical surface as yellow, grey, red, and blue spheres with radii exactly half their van der Waals radii. The valence electron density of the quantum mechanical system is represented by the green contours, which appear only where $n(\vec{r}) = 0.1$ Å$^{-3}$ and $n(\vec{r}) = 0.013$ Å$^{-3}$. The purple-blue background appears where the fluid oxygen density $N_O$ reaches at least half the bulk density of liquid water ($N^{(b)} = 0.033$ Å$^{-3}$) mapped onto a plane with surface normal pointing out of the page. The red and blue contours represent the oxygen and hydrogen densities in the liquid, respectively, and they only appear where $N_O(\vec{r}) =$ $N_H(\vec{r})/2 = 1.1 N^{(b)}$. Thus, contours are only present in locations with significant shell structure in the fluid.

## Surface reactivity characterization

*Electrochemical cell*

Scanning electrochemical microscopy experiments were performed using either an SECM 900 or SECM 920D electrochemical workstation (CH Instruments, Austin, TX). Au and Pt tips for SECM were fabricated using established methods by sealing metal microwires (either Pt 15 µm, Pt 25µm or Au 25µm from Goodfellow Corporation, Oakdale, PA) in capillary tubing and sharpening the resulting microelectrodes.[32]

For cyclic voltammetric, impedance spectroscopy, and SECM measurements, STO samples were placed in a homemade Teflon cell with an o-ring exposing an area of 0.5 cm$^2$ to the electrolyte. Electrical contact to the STO was made using indium wire with copper tape on top of the STO surface outside of the o-ring.

*Surface interrogation SECM*

Surface interrogation SECM (SI-SECM) measurements were performed in a solution of 120 µM K$_3$[Fe(CN)$_6$] in 0.1 M NaOH thoroughly deoxygenated with high purity nitrogen. Before the chronoamperometric SI-SECM experiment, a Pt tip was positioned to a height of ~2 µm above the STO surface as judged from the negative feedback approach curve obtained at the unbiased semiconductor in dark conditions (*i.e.*, no UV irradiation). The SECM tip was poised at 0.05 V vs. Ag/AgCl to reduce ferricyanide [Fe(CN)$_6$]$^{3-}$ to ferrocyanide, [Fe(CN)$_6$]$^{4-}$, and the same electrode reaction was used for SI-SECM transients. At this potential, the Pt tip exhibits a negligible activity towards oxygen reduction, so only mediator molecules are sensed by the tip. The ferri/ferrocyanide redox pair can then be used confidently to redox-titrate photogenerated adsorbed species[32,33] as depicted in Figure s14.

To probe the generation of adsorbed intermediates, the STO was poised under illumination from the optical fiber at various applied potentials for 25 seconds. Then, illumination was shut off and the STO was allowed to rest at open circuit, in that order. The SECM tip was then activated using a chronoamperometric step to 0.05 V vs. Ag/AgCl to generate the ferrocyanide ion which is reactive with reducible adsorbed species at the substrate. An enhanced current is observed if this reaction takes place and as long as reducible species are present at the substrate. Once this reaction proceeds to completion, the enhanced transient decays to the current level of a blank chronoamperometric step in the absence of photogenerated species. Thus, the tip current exhibits a transient positive feedback that decays to a steady state negative feedback. The difference current between surface interrogation chronoamperogram and the blank is integrated and used to yield the photogenerated O intermediate per unit of STO surface.[34]

Typical SI-SECM curves are shown in Figure s15, where the adsorbate charge corresponds to the integrated charge depicted by the yellow area in this figure. A correction to the experimental charge integrated must be introduced to account for the projected area of the mediator's diffusion layer on the extended electrode surface. We introduced this correction in our calculations by performing finite element simulations using the COMSOL Multiphysics 4.2 software for the corresponding diffusion and kinetic problem. We used a similar approach to that previously reported[33,34], where Fickian diffusion in the simulation domain representing the solution is cou-pled to a surface chemical process representing the substrate. Figure s16 shows a summary of the simulation model. Briefly, the tip electrode activates the reducing mediator *R* from originally present *O*. *R* diffuses towards the substrate and titrates surface adsorbate (concentration $\Gamma_O$) with a rate $k_{SI}$. The chemical reaction activates a positive feedback loop that increases the diffusive flux of the reacting mediator species as long as $\Gamma_O$ supports it. The mediator flux at the tip is integrated to yield a current as the output. As shown in the inset of Figure s15 simulated and experimental results are in excellent agreement, where the background subtracted transients overlap almost completely using only the assumption that the rate of reaction at the substrate is diffusion limited.[33,34] These results allow us to confidently extract the surface coverage information presented in Figure s17. Finally, for comparison to the real area of STO, we calculated that the area occupied by one unit cell in STO(100) is approximately 0.16 nm$^2$.[35] This yields a limiting coverage of 100 µC/cm$^2$ for a singly charged O species, e.g. ·OH$_{(ads)}$. Experimentally, we observe that upon saturation of the surface, a limiting coverage of $237 \pm 42$ µC/cm$^2$ is obtained (n=3 samples), roughly equal to two charge equivalents per unit cell, i.e. two ·OH$_{(ads)}$ or one O$_{(ads)}$. Regardless of its exact nature, it is interesting to notice that at potentials closer to the open circuit of the STO electrode, the quantity of this adsorbate decreases. This could be due to incomplete electrode activation or coexistence with other plausible intermediates, for example, H$_{(ads)}$, which could be required to sustain H$_2$ production at open circuit on the STO crystal.

*Ex situ surface analysis*

Atomic force microscopy (Scanning Probe Microscope D3100, Veeco), and X-ray photoelectron spectroscopy (SSX-100, Surface Science Instrument) were measured on the sample before and after applying potential to the STO. AFM images show that the surface morphology of the sample after the surface preparation. The feature of terrace is clearly observed. Even after the water splitting reaction, the terraces are still intact, and the surfaces are not rough. XPS analysis shown in Figure s18 for Sr, Ti, and O did not change considerably in oxidation state or elemental composition. However, we need to take account that the electron mean free path at this X-ray energy (Al K$_\alpha$, hυ=1486eV) would be 10 nm. If the surface change occurred only in the first or two monolayers, XPS may not be sensitive to a surface change.

*SECM oxygen collection experiments*

The water splitting reaction, Equation S4, produces hydrogen and oxygen when the STO surface is illuminated with UV light at open circuit. With the intention of obtaining a rapid and versatile way of assessing reactive changes in our samples just after electrochemical activation, we used SECM in the substrate generation – tip collection mode (SG/TC).[36] Here, an SECM tip can be used to collect the products of reaction from a localized area of the electrode (approximately the size of the tip microdisk). Multiple sites can be evaluated by performing approach curves on different places in the sample. Since the SECM tip is placed very close to the STO electrode, we can detect changes in the product output upon illumination almost instantaneously (~ 0.8 ms here) which allows for comparing signal to background during light chopping measurements. Likewise, we can measure changes in the product output of our samples before and after oxidative excursion under UV light in the same setup immediately after these changes have taken place.

We focused on the localized collection of oxygen produced at the STO substrate, for which the oxygen reduction reaction (ORR, Equation S5) in basic medium was used to detect this species.

$$2H_2O \rightleftharpoons 2H_2 + O_2 \quad (S4)$$
$$O_2 + 2H_2O + 4\bar{e} \rightleftharpoons 4OH^- \quad (S5)$$

$E^0 = 0.201$ vs. Ag/AgCl

Au/Hg amalgam microelectrodes were to detect oxygen using the ORR without interference from other species produced during the water splitting reaction, e.g. $H_2$. ORR at Hg in base is convenient since it can be fully activated to yield the 4 electron reaction, Equation S5, without the generation of intermediates in solution (e.g. $H_2O_2$).[36]

Au/Hg electrodes [37] were prepared by immersing a 12.5 μm Au SECM tip in a solution containing 5 mM $Hg(OAc)_2$ in 0.1 M $KNO_3$ with 0.5% $HNO_3$. A reduction step was applied in order to deposit Hg at diffusion limited rate for 5 minutes. This produced a micro-hemisphere that was gently blotted with a piece of paper to expose the amalgamated Au/Hg microdisk. After thorough rinsing, the electrodes were used to approach the STO surface at open circuit.

Oxygen collection experiments with the Au/Hg tips were performed in 0.1 M NaOH, unless indicated otherwise and using an ozone-free Xe lamp (6258, Newport Corporation). The SECM tips were consistently approached to c.a. 7±1 μm from the STO surface using the ORR as a negative feedback probe. Once positioned, the solution was thoroughly deoxygenated using high-purity Ar. For a given run, the tip potential was held at a potential where the four-electron ORR occurred (e.g. -1.4 V vs. Ag/AgCl) while the STO sample rested at open circuit and completely disconnected from any other electronic element in the electrochemical cell. Cycles of illumination and darkness, i.e. chopping, were applied to produce results such as the ones shown in the main text. The samples then underwent an oxidative treatment to potentials near 1.0 V vs. Ag/AgCl. We then repeated the illumination/darkness cycles explained above, with the STO resting at open circuit once more. It is important to emphasize that measurements with the original and modified STO surface were performed to the best of our experimental effort under exactly the same conditions, i.e. same illumination levels, electrolyte composition and other environmental effects. Figure s19 shows a replicate experiment where an increase similar to that shown in the main text is observed. We observed that an increase in the activity can be obtained even after one single potential excursion to 1.0 V vs. Ag/AgCl at 0.01 V/s; however for the purpose of comparison, we typically applied 1.0 V vs. Ag/AgCl for periods of 20 minutes. After this time, the activity does not seem to increase further. This sample activity increased by a factor of 2.6 ± 0.9. This value was obtained after analyzing five different STO samples in at least two different sites.

Control experiments were run to verify that the $O_2$ collection transients represented the photoactivation of STO. When the output from the Xe lamp was filtered with a 400 nm cutoff filter while retaining exactly the same configuration in the setup, no increase in collection current was observed (i.e. no change with respect to the background). This is expected, since our STO samples are practically inactive when irradiated with the visible portion of the spectrum, but the control also precludes the possibility that the increase in activity is an artifact of thermal effects, either at STO or at the Au/Hg electrode. Likewise, when the tip electrode is raised to a distance of ~200 μm above the STO surface only a small (~ few pA) and delayed increase in SECM tip cur-rent is observed. This implies that the transient effects are not a product of photoactivation of any process in the bulk of the electrolyte and that as expected, when we approach closer to the STO substrate, the collection efficiency increases.[36]

The same STO surface that displays activity in NaOH is not measurably active in 0.1 M $H_2SO_4$ at open circuit, per the SECM experiments shown in the main text. When short circuited to a Pt electrode, STO in 0.1 M $H_2SO_4$ does show however a comparable activity to that in 0.1 M NaOH (i.e. with hydrogen evolution at the Pt counter electrode). This suggests that while STO in acid is able still to inject electrons into its conduction band and transport them to a short-circuited cathode to perform the assisted water splitting reaction, it cannot split water on its own surface.

$O_2$ collection currents obtained with the SECM setup can be translated into more familiar measures of product flux. We performed semi-quantitative calculations using the finite element method provided by the COMSOL Multiphysics v. 4.2 software to estimate the equivalence of the measured tip currents to the product flux of $O_2$ from the STO surface. The simulation space was modeled using the distances and dimensions of the electrodes described above. A diffusion coefficient of $2.5 \times 10^{-9}$ $m^2/s$ was used for $O_2$ in 0.1 M NaOH.[38] Figure s20 shows an approximated relationship between typical tip currents obtained in our experiments and the flux of $O_2$ from the STO surface. The tip currents were obtained from the differential current $i_{activation} - i_{background}$. From these results, we obtain that 100 pA of measured ORR current is equal to 210 $\mu mol\, h^{-1} m^{-2}$ of $O_2$ evolved. Inspection of Figure 2 in the main text and Figure s19 show that this is roughly what is produced at the open circuit STO after oxidative treatment. With an illumination power of 15 $mW/cm^2$ (obtained after using an optical fiber), and assuming an area of 0.16 $nm^2$ per STO unit cell we can estimate the turn over frequency of our STO catalysts. The estimated turn over frequency for oxygen evolution for the open circuit STO is 0.006 $O_2$ $site^{-1}s^{-1}$, 0.06 $O_2$ $site^{-1}s^{-1}$ for a STO crystal at open circuit but short-circuited STO to a Pt counter electrode and 0.17 $O_2$ $site^{-1}s^{-1}$ for an STO crystal biased to 1.0 V vs. Ag/AgCl. These numbers are expected to change as a function of illumination intensity but provide a basis for activity comparisons.

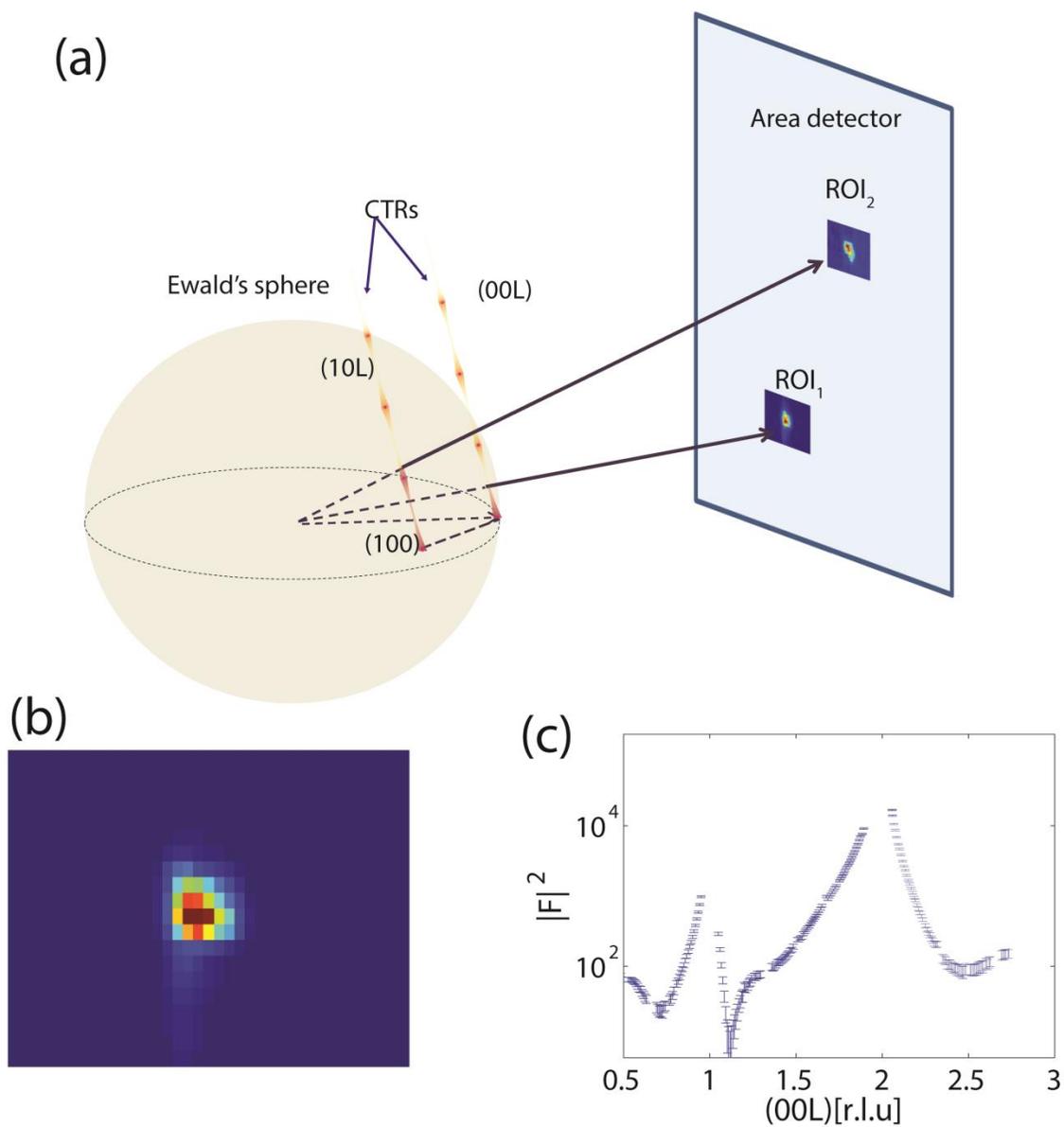

**Figure S1.** **(a)** Schematic diagram showing how multiple CTRs are measured simultaneously with the large area detector. **(b)** Diffraction image of (00L) CTR spot. L value for this particular CTR spot is 1.75. **(c)** Complete $|F|^2$ vs. (00L) for fresh, *n*-doped STO in air.

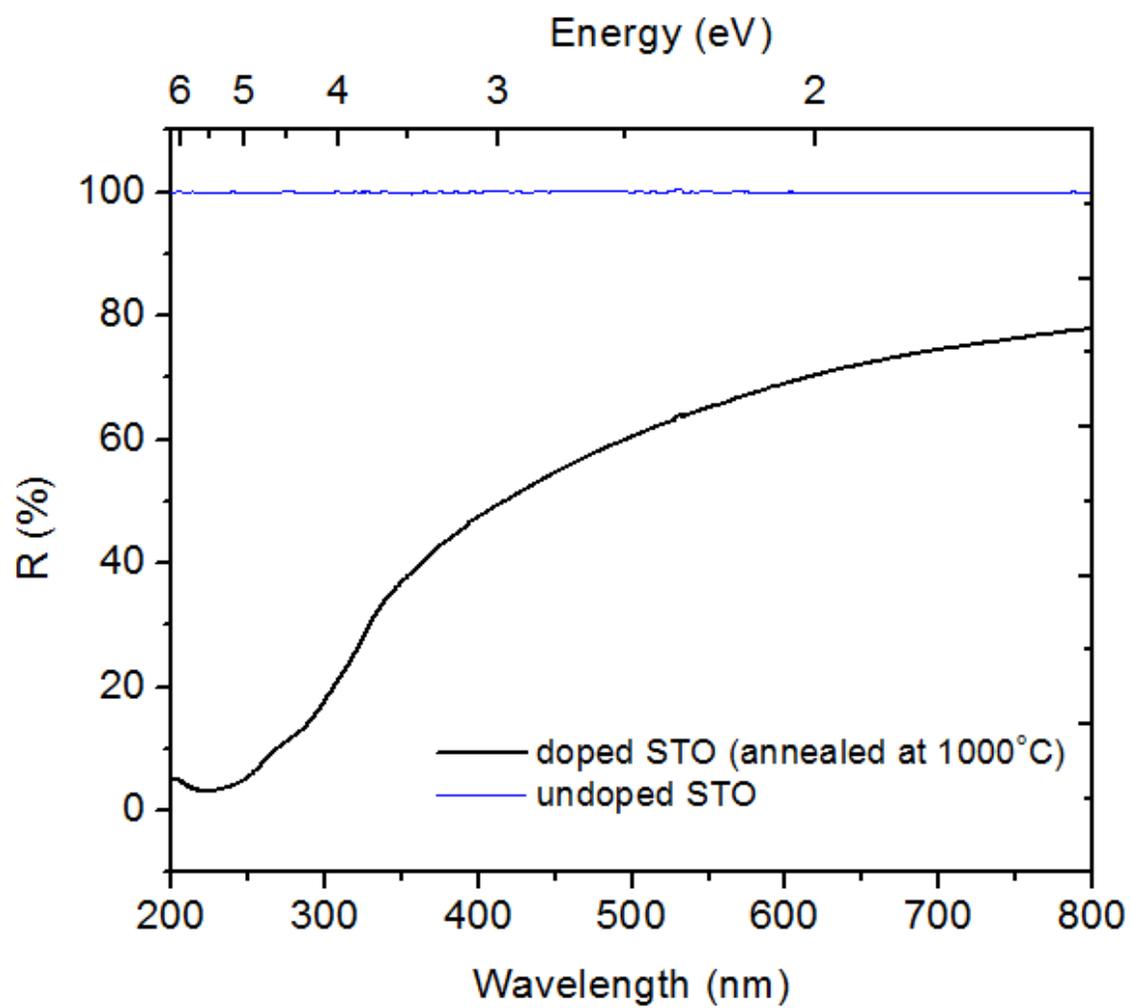

**Figure S2**. UV-Vis reflectance of oxygen vacancy n-doped STO substrates versus an undoped reference.

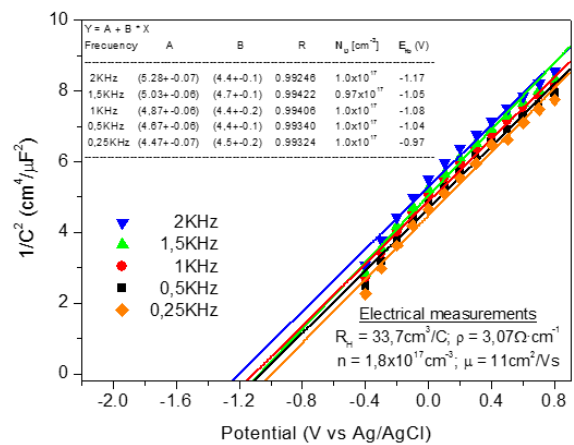

**Figure S3**. Mott-Schottky of n-doped STO electrodes used in this study. Electrolyte: 0.1 M NaOH and in the dark.

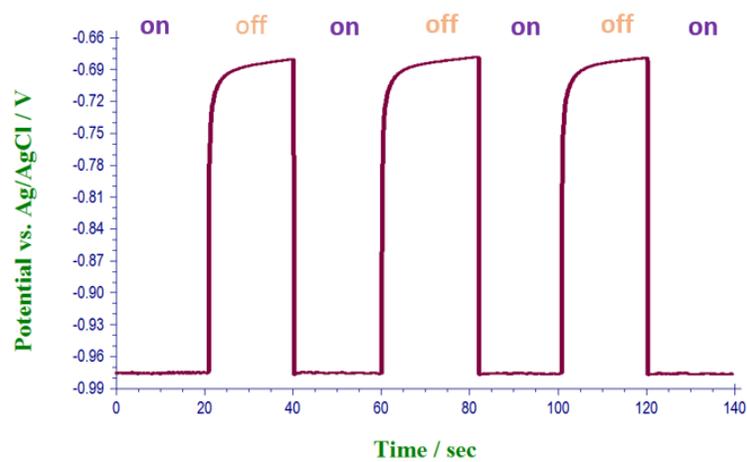

**Figure S4**. Typical open circuit potential transients for n-doped STO electrodes in 0.1 M NaOH. Top labels indicate Xe lamp status.

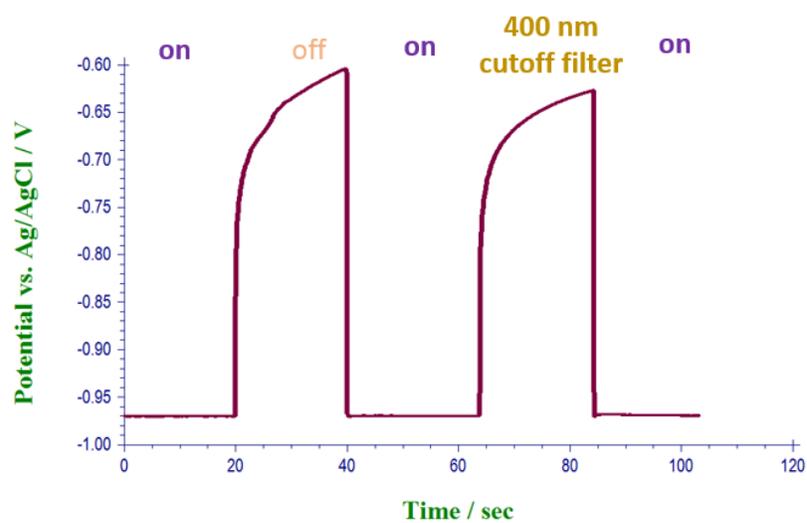

**Figure S5**. Typical open circuit potential transients for n-doped STO electrodes in 0.1 M NaOH with a 400 nm cutoff filter. Top labels indicate Xe lamp status.

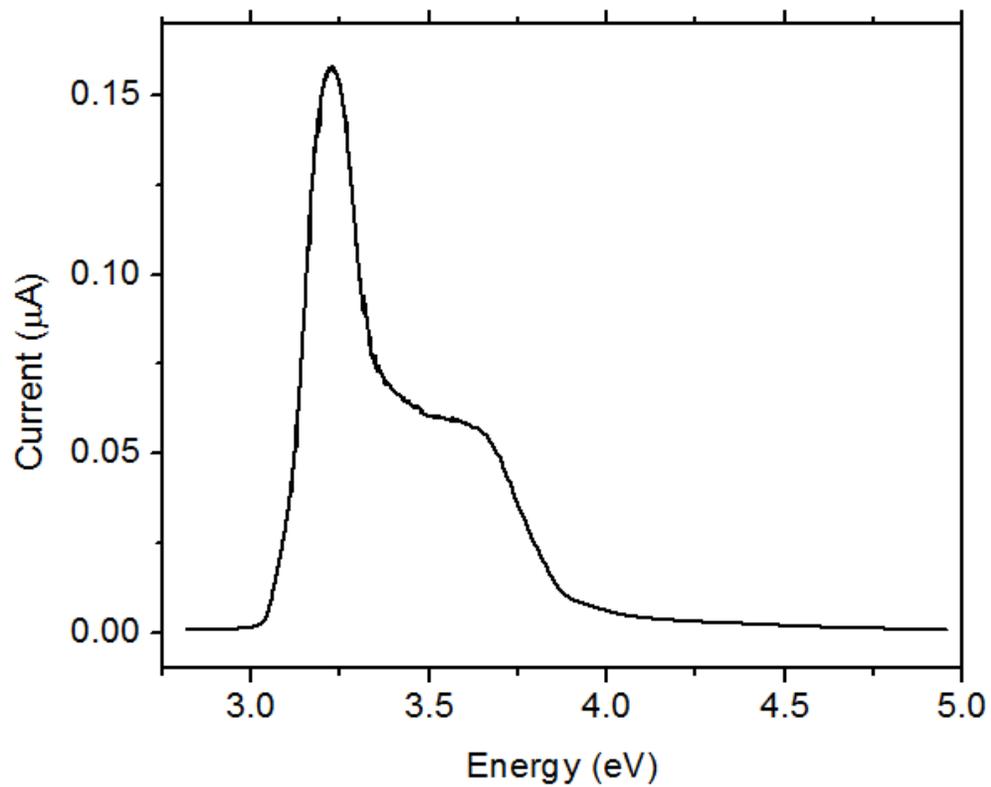

**Figure S6**. Photoaction spectrum of n-doped STO. Electrode area ~ 0.5 cm$^2$, Electrolyte: 0.1 M NaOH. STO electrode was poised at +0.8 V vs. Ag/AgCl. Sample was illuminated with a 200 W Hg(Xe) lamp (±20 nm spectral resolution).

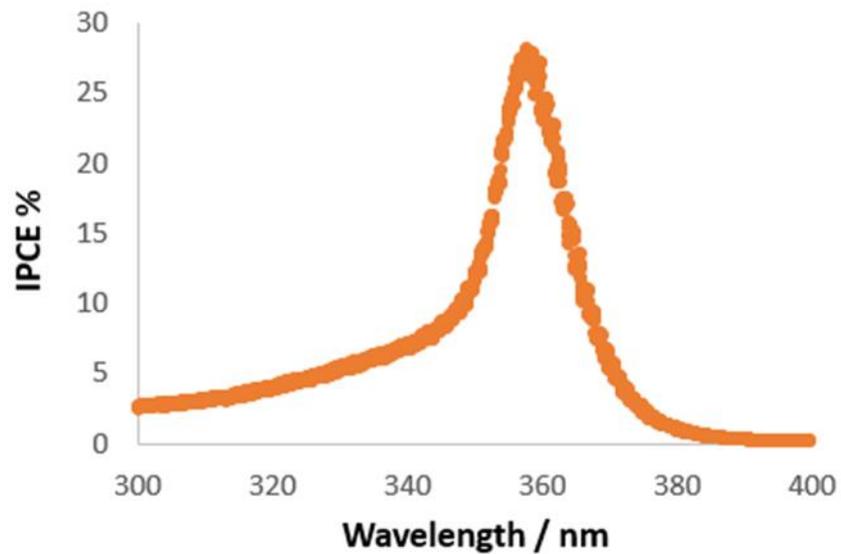

Figure S7 Incident photon conversion efficiency for a typical strontium titanate sample with oxygen vacancies on the order of $10^{17}$ cm$^{-3}$

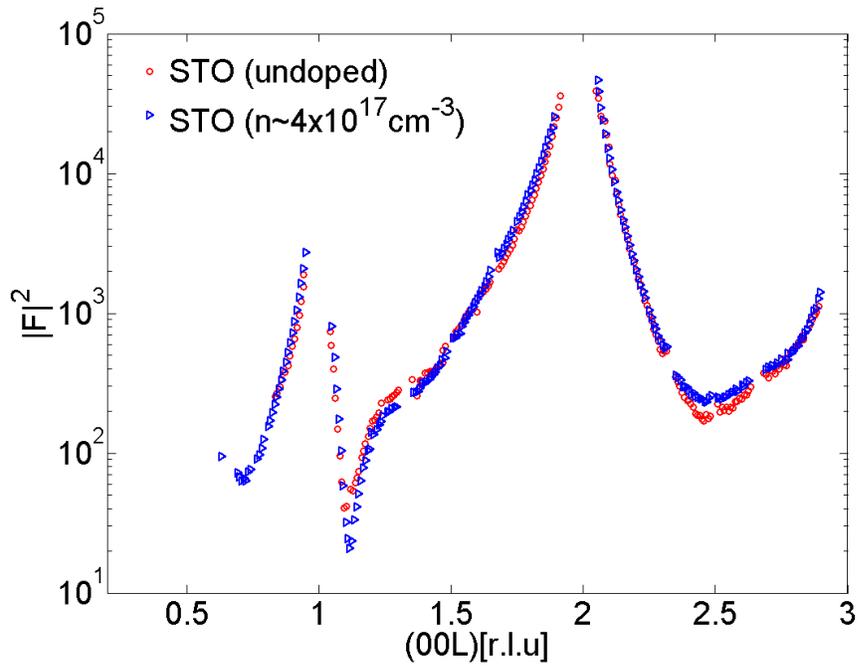

**Figure s8.** CTR of STO with and without doping. Sample is in air.

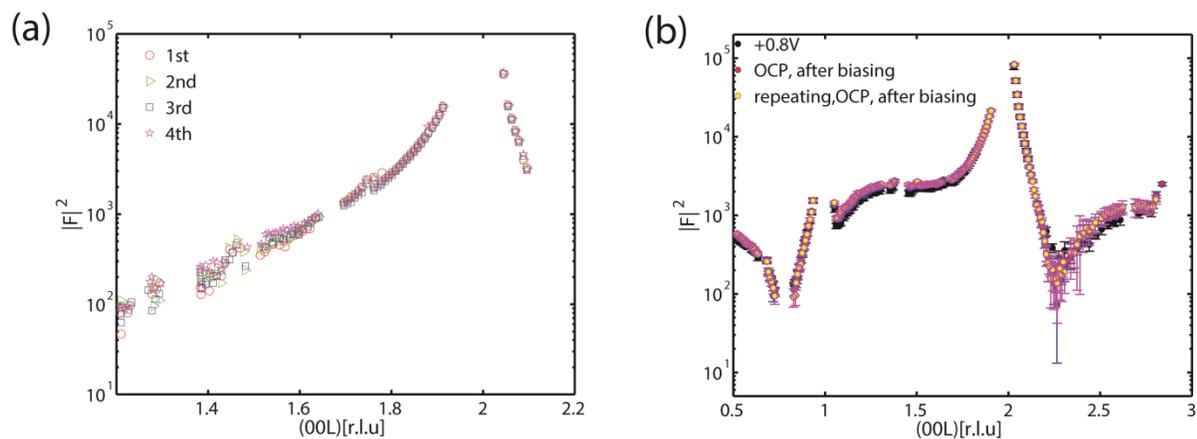

**Figure s9.** Demonstration of stability. **(a)** Repeated CTR measurements over 10 hours on the same STO sample at open circuit potential. **(b)** CTRs at **(1)** +0.8V vs. Ag/AgCl, **(2)** after returning to open circuit potential, and **(3)** repeating the measurement at open circuit potential.

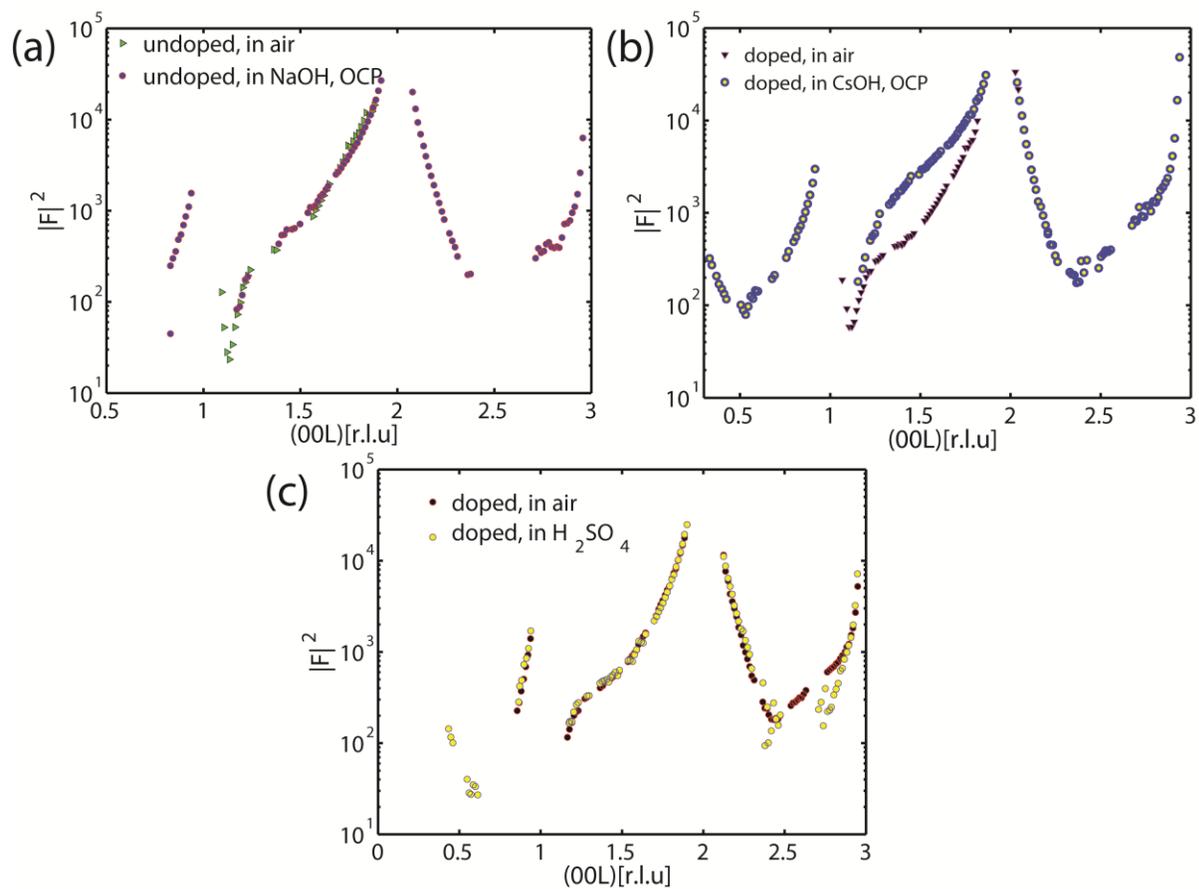

**Figure s10.** CTR measurements of **(a)** undoped STO in air and in 0.1 M NaOH; **(b)** doped STO in air and in 0.1 M CsOH at open circuit; **(c)** doped STO in air and in 0.1 M $H_2SO_4$ at open circuit.

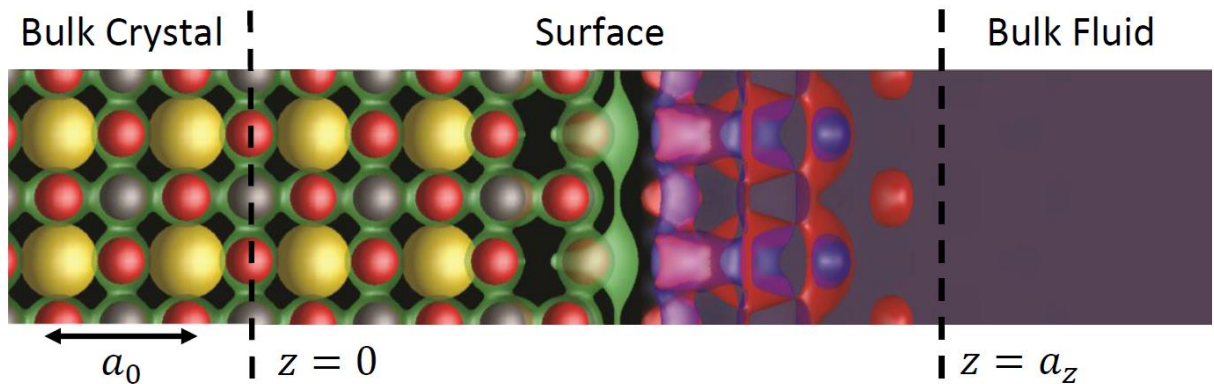

**Figure s11** Schematic of X-ray structure calculation from first principles, with a region representing each of the 3 terms from Equation S3. The DFT lattice constant is $a_0$ and $a_z$ is half the length of the DFT supercell in the z direction (see Figure s12)

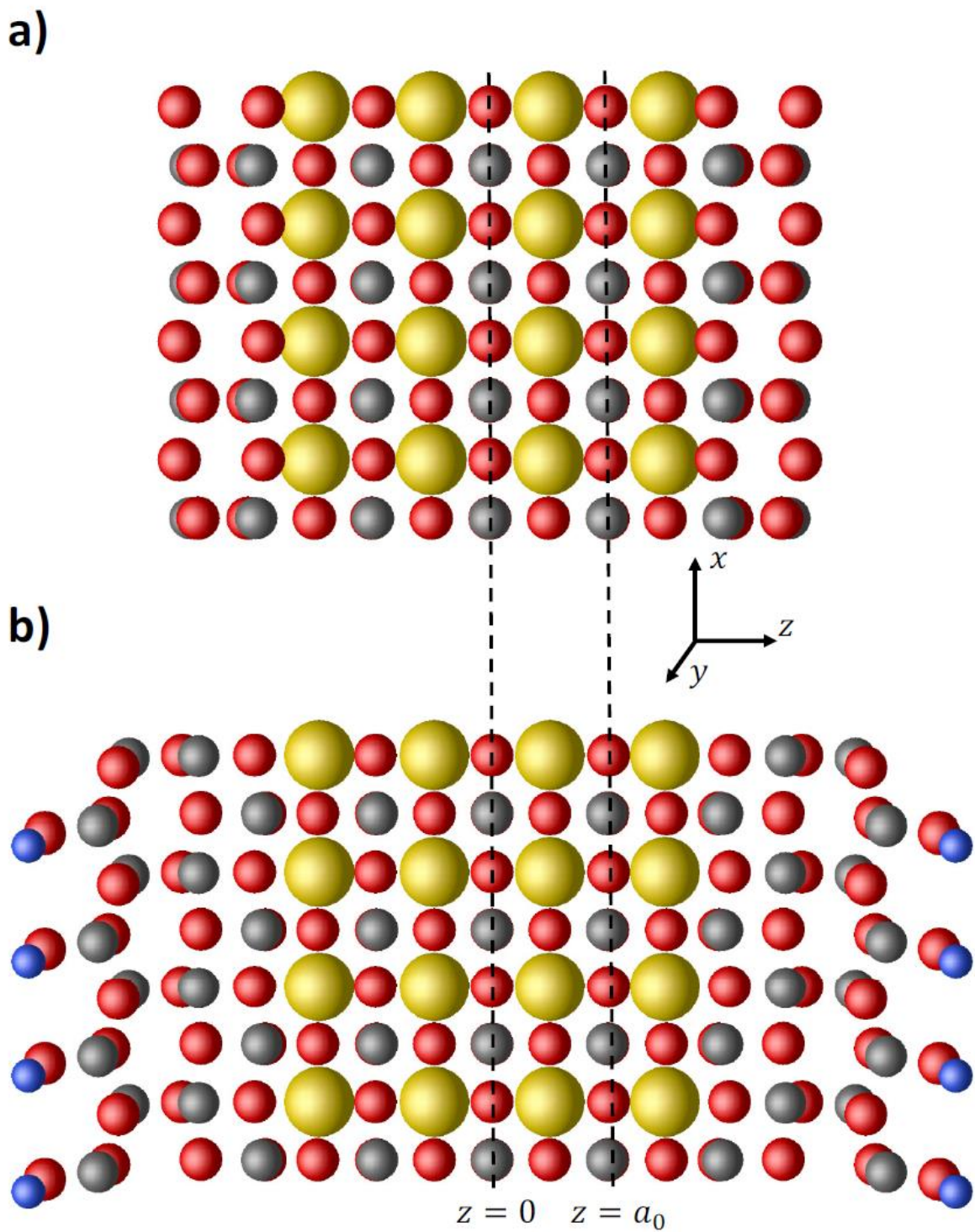

**Figure s12** The k-point sampled DFT supercells for the a) untrained and b) trained surfaces. Sr, Ti,O, and H atoms are yellow, silver,red, and blue respectively. Unit cells are truncated in the z-direction and should extend to $z = \pm a_z$, where $a_z \approx 20a_0$. The trained surface has been rotated 90° in the x-y plane from Figure 3(d) to better display the anatase-like stacking of the Ti atoms.

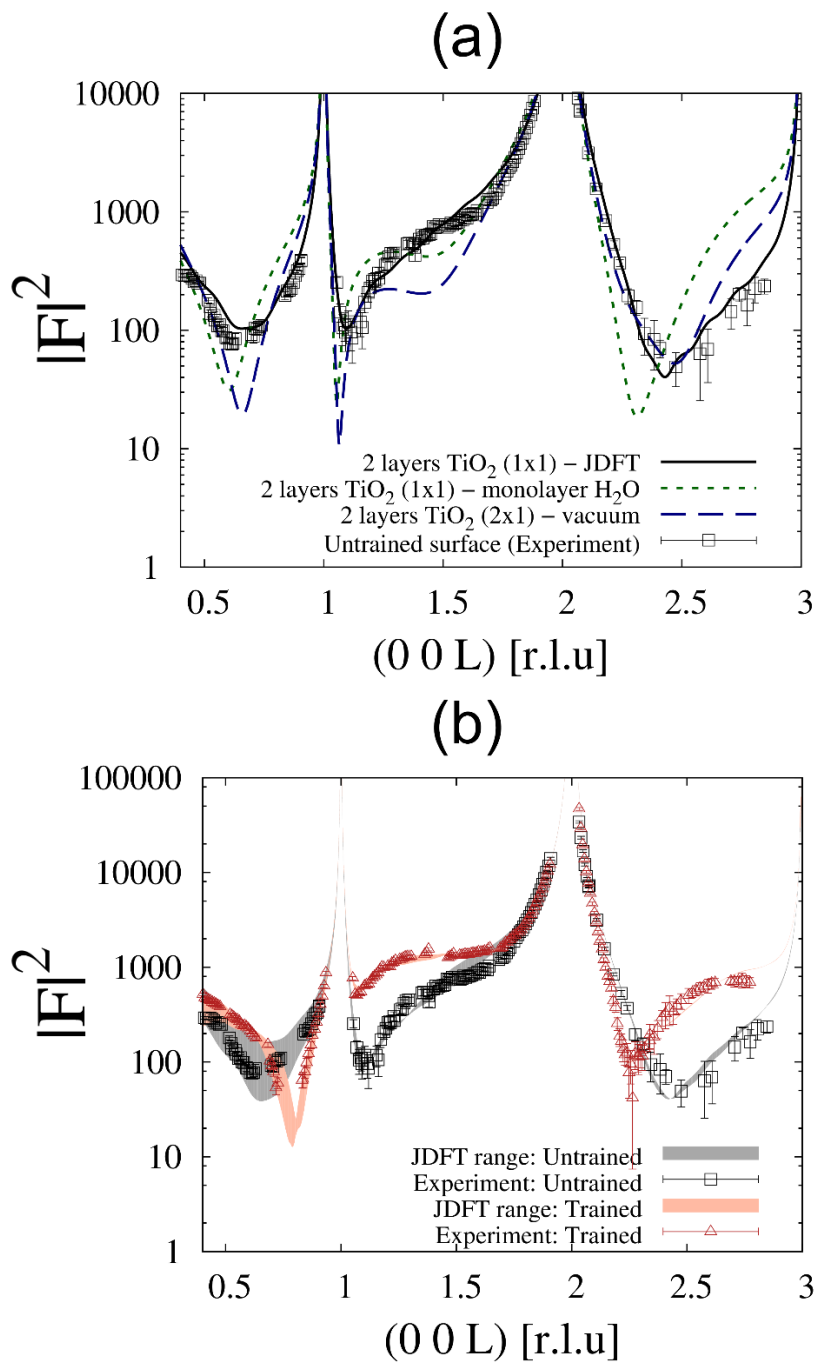

**Figure s13** (a) JDFT predicted CTR (black solid line) which agrees best with X-ray measurements (black squares) for untrained surface compared to the predicted CTR for the same surface composition with a single water molecule per unit cell relaxed within vacuum DFT (dotted green line) and the CTR of the 2x1 surface reconstruction[19] which has the minimum energy in vacuum DFT (dashed blue line). (b) JDFT results from Figures 3(a) and (c) with sensitivity analysis performed on the construction of the coupling functional (describing interactions between the fluid and the electrode). Black and red points represent the experimental data for the untrained and trained surfaces respectively, while the corresponding grey and pink shaded regions represent the range of physical predictions for the effect of a non-reacting fluid on the surface.

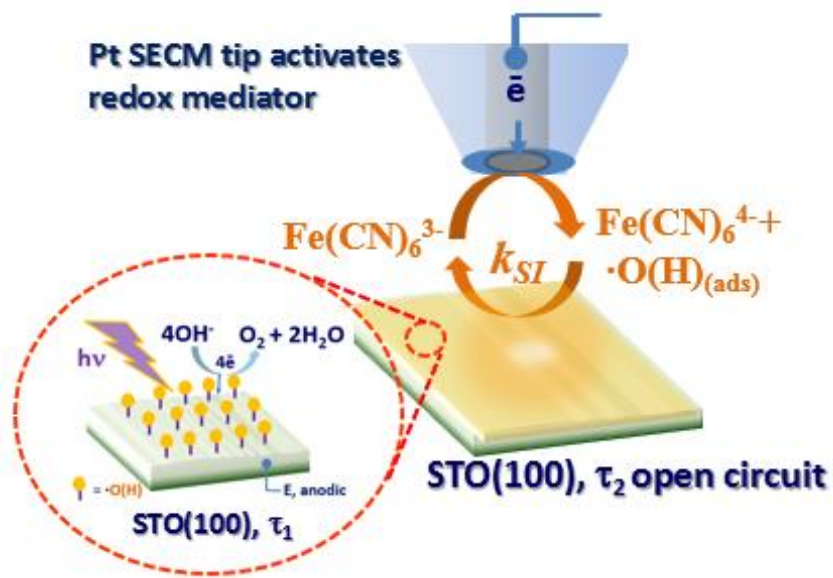

**Figure s14**. Schematic of SI-SECM for the interrogation of photogenerated oxygen on strontium titanate using ferri/ferrocyanide as redox mediator

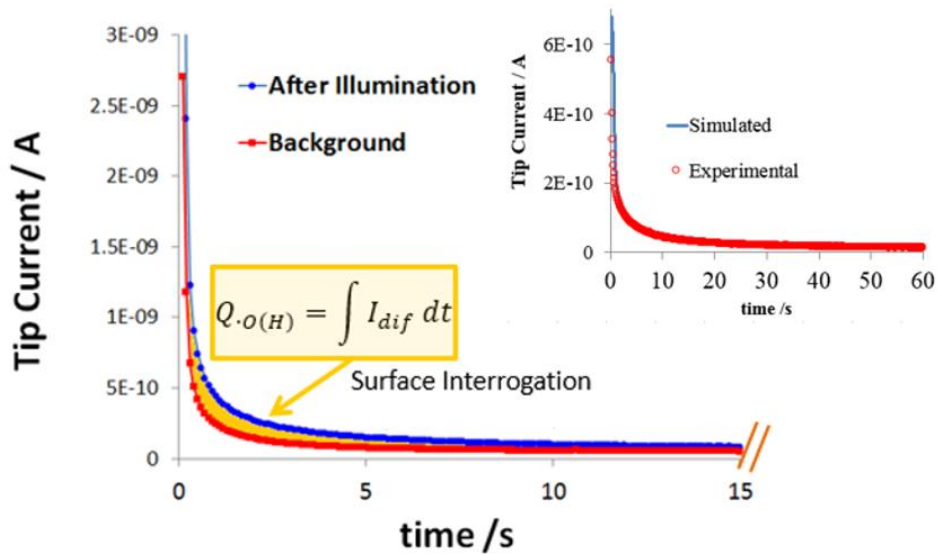

**Figure s15**. Surface interrogation of adsorbed, photoelectrochemically generated O(H) intermediate at the STO surface using tip generated $[Fe(CN)_6]^{4-}$ as redox mediator. Pt tip: radius 7.5 μm at $d = 2$ μm from STO. Mediator 120 μM $K_3[Fe(CN)_6]$ in 0.1 M NaOH. Illumination using a 200 W Hg(Xe) lamp. Experimental result shows a typical SI-SECM transient at full coverage obtained after biasing STO to 0.8 V vs. Ag/AgCl and compared to the background transient obtained in the dark after applying a reducing step to -0.9 V vs. Ag/AgCl. Inset shows the comparison of the background subtracted transients for the same experimental result and the simulated case, as explained in text. Simulated adsorbate charge: $Q = 240$ μC/cm$^2$.

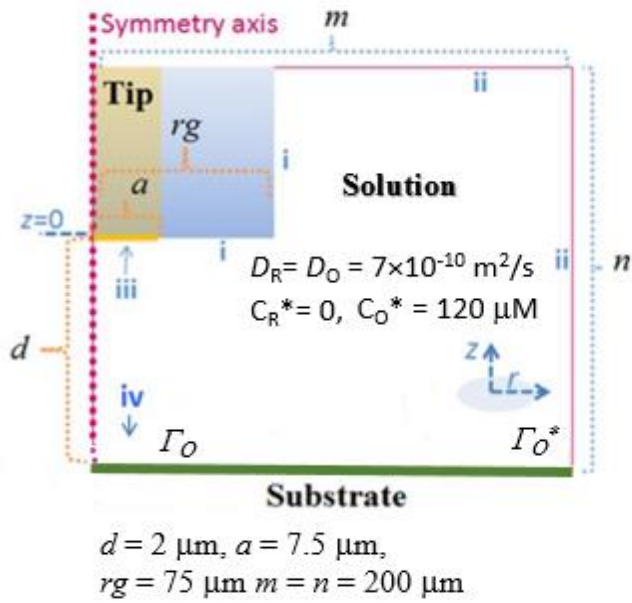

**Figure s16**. Summary of simulation model for SI-SECM.

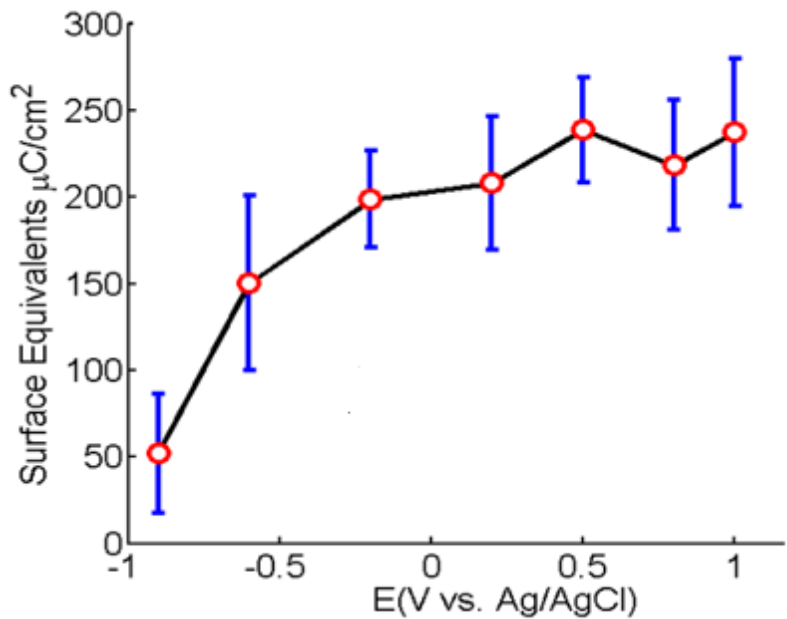

Figure s17. Equivalent charge for photogenerated oxygen on $SrTiO_3$ electrodes after illumination in NaOH using SI-SECM.

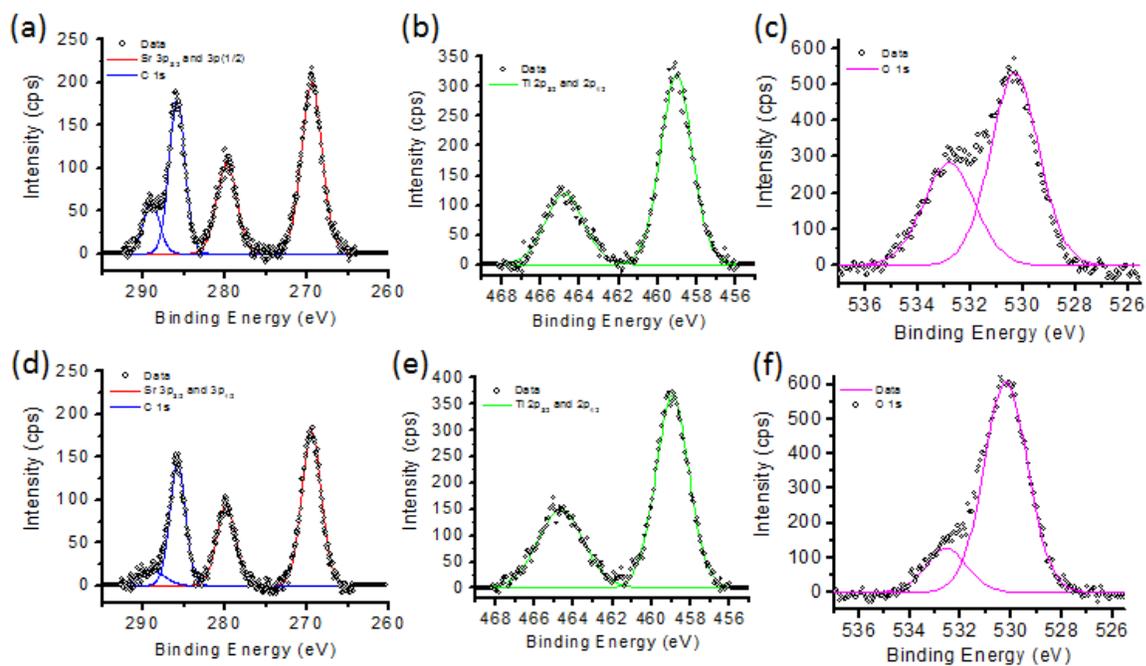

**Figure s18.** XPS measurement of the STO sample before (a)-(c) and after (d)-(f) photo-assisted water splitting. (a) and (d) are the Sr 3p electron state; (b) and (e) are the Ti 2p state; (c) and (f) are O 1s state.

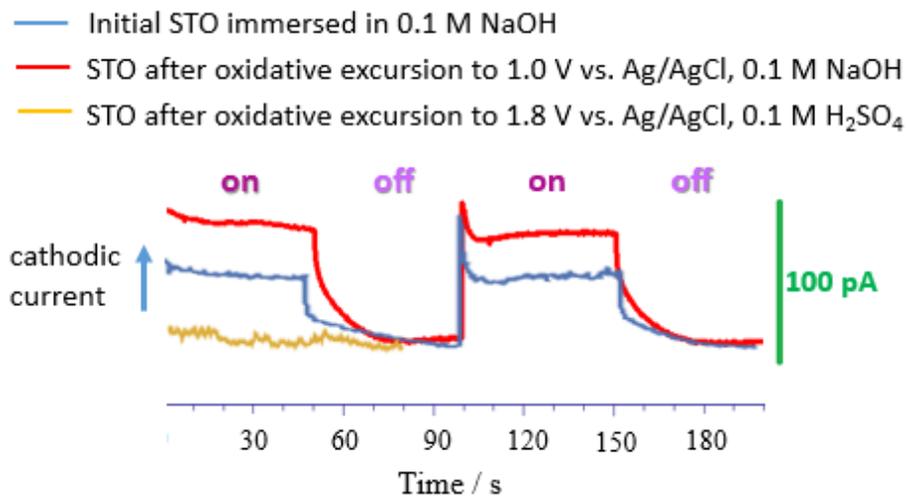

**Figure s19.** $O_2$ collection transients at Au/Hg SECM tip (radius = 12.5 μm at 7 μm from substrate) during photoactivation of STO at open circuit. Lamp status is indicated on top of curves. After oxidative treatment in 0.1 M NaOH, there is an increase in the $O_2$ generation of the STO electrode, while in acidic conditions, no photoresponse is observed even after oxidative treatment. These are complementary results to those in Figure 2(d) in the main text.

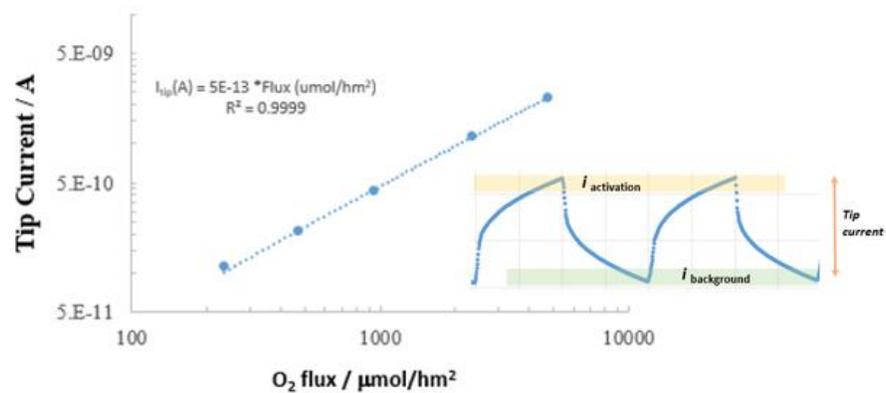

**Figure s20**. Simulated relationship between $O_2$ flux from substrate electrode and tip current response under the conditions of Figures S19 and Figure 2(d) in the main text. Tip current was estimated as $i_{activation}$ - $i_{background}$ as shown schematically in the inset.

| atom | $z/a_0$ |
|------|---------|
| Ti   | 0.00000 |
| O    | 0.00000 |
| O    | 0.00000 |
| Sr   | 0.50000 |
| O    | 0.50000 |
| O    | 0.99069 |
| Ti   | 0.99874 |
| O    | 1.00812 |
| O    | 1.49831 |
| Sr   | 1.49964 |
| O    | 1.93530 |
| Ti   | 1.99698 |
| O    | 2.06580 |
| O    | 2.49304 |
| Ti   | 2.55719 |
| O    | 2.65270 |

**Table s1**. JDFT-calculated atomic positions for the immersed sample in the $z$-direction (normal to the surface) in units of lattice constant $a_0$.

| Atom | $z/a_0$ | $(z - z_{DFT})/a_0$ | occupancy | Debye-Waller $\sigma/a_0$ |
|---|---|---|---|---|
| Ti | 0.00000 | 0.000 | 1 | 0.004 |
| O | 0.00000 | 0.000 | 1 | 0.005 |
| O | 0.00000 | 0.000 | 1 | 0.005 |
| Sr | 0.50000 | 0.000 | 1 | 0.004 |
| O | 0.50000 | 0.000 | 1 | 0.005 |
| Ti | 1.00719 | 0.014 | 1 | 0.004 |
| O | 0.94512 | 0.057 | 1 | 0.086 |
| O | 0.99236 | 0.018 | 1 | 0.005 |
| Sr | 1.51373 | 0.017 | 1 | 0.004 |
| O | 1.51281 | 0.003 | 1 | 0.169 |
| O | 1.96606 | 0.002 | 1 | 0.026 |
| Ti | 1.99768 | 0.004 | 1 | 0.071 |
| O | 2.07735 | 0.022 | 1 | 0.190 |
| O | 2.56112 | 0.040 | 1 | 0.005 |
| Ti | 2.58703 | 0.046 | 1 | 0.004 |
| O | 2.64567 | 0.037 | 1 | 0.005 |
| Ti | 3.12121 | 0.025 | 1 | 0.228 |
| O | 3.22553 | 0.011 | 1 | 0.028 |
| O | 3.31986 | 0.002 | 1 | 0.059 |
| Ti | 3.42910 | 0.018 | 0.4761 | 0.004 |
| O | 3.91892 | 0.047 | 0.9999 | 0.095 |
| H | 4.01449 | 0.000 | 1 | 0.000 |

**Table s2**. Best fit for the activated sample: atomic positions $z$ and discrepancy from DFT positions $z - z_{DFT}$ (normal to the surface), best fit occupancies, and best fit Debye-Waller rms displacements. All positions are in units of lattice constant $a_0$.